# Surface Chemistry-based Continuous Separation of Colloidal Particles via Diffusiophoresis and Diffusioosmosis

Adnan Chakra[a,b,c], Christina Puijk[c], Goran T. Vladisavljević[b], Cécile Cottin-Bizonne[d], Christophe Pirat[d], and Guido Bolognesi[c,✉]

[a]Department of Chemistry, Imperial College London, London, W12 7TA, United Kingdom
[b]Department of Chemical Engineering, Loughborough University, Loughborough, LE11 3TU, United Kingdom
[c]Department of Chemistry, University College London, London, WCH1 0AJ, United Kingdom
[d]Institut Lumière Matière, UMR5306 Université Claude Bernard Lyon 1, 69622, Villeurbanne, France

**The separation of colloidal particles based solely on their surface properties is a highly challenging task. This study demonstrates that diffusiophoresis and diffusioosmosis enable the continuous separation of carboxylate polystyrene particles with similar sizes and zeta potentials but distinct surface concentrations of carboxyl groups. The particles are exposed to salt concentration gradients generated in a double-junction microfluidic device. Through experimental and theoretical analyses, we demonstrate how the particle dynamics are influenced by their zeta potential sensitivity to the local salt concentration, which in turn is affected by surface conductance effects induced by the surface carboxyl groups. Consequently, colloids with comparable zeta potentials but differing surface concentrations of carboxyl groups can be separated with 100% efficiency. This approach, which employs a simple, easy-to-operate device, has discipline-spanning potential for the continuous separation of colloids distinguished solely by surface properties that influence their zeta potential sensitivities, like roughness, permeability, heterogeneity, and chemical composition.**

colloids | microfluidics | particle separation | diffusiophoresis | diffusioosmosis | zeta potential | surface conductance
Correspondence: *g.bolognesi@ucl.ac.uk*

## Introduction

The separation of nano- and micro-particles based on their physical and chemical properties is of great importance in a wide range of applications, including diagnostics (1), drug delivery (2), chemical and biological analyses (3), food processing (4), biofuel production (5), pharmaceutical industry (6), and environmental monitoring and remediation (7). Microfluidic-based particle separation methods offer several advantages compared to conventional bulk approaches (8). Higher precision, smaller footprint, improved energy efficiency, and reduced consumption of costly or hazardous samples and reagents make microfluidic methods a more efficient, cost-effective, sustainable, and safer alternative. These methods are particularly advantageous for the rapid and continuous processing of small volumes of materials, such as biofluids or expensive nanomaterials.

Active microfluidic separation techniques rely on external fields and auxiliary equipment and are most effective in terms of high separation efficiency and throughput (9). Conversely, passive hydrodynamic techniques rely on interactions between particles, flows and channel structures; they can easily operate with simpler experimental set-ups, but they are only adequate for separation based on particle size and elasticity (10, 11). Alternatively, electrolyte diffusiophoresis and diffusioosmosis – namely, the spontaneous relative motion between adjacent liquid and solid phases induced by a solute concentration gradient – have been successfully applied for the passive and continuous separation of colloids in microfluidic devices (12–18). Since the diffusiophoresis mobility of colloids depends on particle size and zeta potential, diffusiophoresis/osmosis-assisted microfluidic strategies have enabled the continuous separation based on particle size and surface charge properties. However, the small magnitude of the diffusiophoresis and diffusioosmosis velocities (1-10 $\mu$m s$^{-1}$) combined to the typically short residence times of the particles in the microfluidic devices (1-10 s) have implied that up to now only particles with significantly different size (i.e., at least one order of magnitude) (17, 18) or zeta potential (i.e., oppositely charged surface groups) (13, 14, 16), could be separated efficiently under continuous flow settings.

The zeta potential $\zeta$ of a colloid or a solid wall depends on the physical and chemical properties of its surface and surrounding electrolyte solution. For particle and wall materials commonly used in microfluidics (e.g., silica, silicon, and polymers), the zeta potential varies with the ionic strength, and thus, electrolyte concentration of the solution (19, 20). This relationship is affected by the solution pH and the surface physico-chemical properties of the solid substrate, including type and concentration of surface ionizable groups, ion adsorption affinity, roughness, and permeability to ions and liquid. In diffusiophoresis/osmosis microfluidic experiments, the difference in ionic strength between the lowest and highest concentration streams is typically of two or three orders of magnitude, potentially leading up to 10-fold variation of particle and wall zeta potential between the two streams (21). Therefore, at least in principle, the continuous separation by diffusiophoresis/osmosis effects could be extended to particles that exhibit similar values of zeta potentials $\zeta(c)$ at a given salt concentration $c$, but different zeta potential sensitivity to varying salinity levels (i.e., $\partial \zeta/\partial c$). Indeed, as the particles are exposed to local environments with varying ionic strengths, their zeta potential can also vary to a different extent, depending on $\partial\zeta/\partial c$, thus leading to different particle dynamics. If possible, this would enable the separation of particles with similar zeta potentials at a given salt concentration but distinct surface physico-chemical properties that influence the relationship between zeta potential and local salt concentration. In practice, the dependence of zeta potential on salt concentration is often neglected, and the assumption $\partial\zeta/\partial c$=0 is commonly used in studies on diffusiophoresis/osmosis effects (12–18, 22–29). This is because constant potential models facilitate the theoretical work and also provides a good quantitative agreement with experimental observations in most cases (21), apart a few exceptions (30–32). As a result, the anticipated opportunity for passive continuous separation of similarly charged particles with different surface physico-chemical properties seems to be out of reach. In a previous study, we presented a physical mechanism whereby diffusiophoresis and diffusioosmosis are closely intertwined, leading to a strong transverse focusing of particles towards the central region of a straight channel under continuous and steady flow conditions. A double $\Psi$-junction device, generating salt concentration gradients from the merging of 0.1 mM and 10 mM LiCl aqueous streams, was developed to exploit this focusing mechanism for preconcentration, sorting, and characterization of liposomes and charged polystyrene beads with a diameter ranging from 20 nm to 1 $\mu$m (18). Again, under the examined conditions, a constant zeta potential model allows the quantitative prediction of the experimental results. Also, charge-based separation was only possible for oppositely charged particles, whereas size-based separation was achieved efficiently only between 20 nm and 1 $\mu$m particles. In this study, we investigated the particle dynamics in the same device, but under varying ionic strengths, ranging from 0.01 mM to 100 mM. Remarkably, we found that in this wider range of salt concentrations, the effect of ionic strength on particle and wall zeta potential becomes apparent, and the experimental observations can be interpreted only by accounting for a salt concentration dependent zeta potential. More surprisingly, we discovered that similarly-sized carboxylate polystyrene particles exhibiting comparable values of zeta potential but different surface concentration of carboxyl-groups and thus opposite zeta potential sensitivities to the local salt concentration, namely $\partial\zeta/\partial c > 0$ and $\partial\zeta/\partial c < 0$, can be continuously separated with 100% separation efficiency. Our findings, supported by experimental and numerical analyses, open up unprecedented opportunities for the diffusiophoresis/osmosis-assisted microfluidic continuous separation of synthetic or biological particles, including macromolecules (33), liposomes (24), exosomes (29), viruses (25) and bacterial cells (34), based exclusively on their surface physico-chemical properties.

## Results

The double $\Psi$-junction micro-chip used for particle focusing in continuous flow settings (18) is illustrated in Figure 1A. The device consists of a narrower (upstream) junction, where the inner inlet channel meets the middle inlet channels and merges into the middle channel of width $w_m$. This is followed by a wider (downstream) junction, where the middle channel meets the outer inlet channels and merges into the main channel (see Supplementary Text for channel sizes). The flow rates of the inner, middle and outer streams were fixed to $3.65\,\mu\mathrm{L\,min}^{-1}$, and the particle dynamics was observed using an epi-fluorescent microscope. Lithium chloride (LiCl) salt was employed to establish solute gradients perpendicular to the flow, by injecting low-concentration LiCl solution ($c_L$) in the inner and middle inlets and either low or high-concentration LiCl solution ($c_H$) in the outer inlets. Colloidal particles were dispersed in the middle inlet stream only. Figure 1 shows the typical dynamics of sub-micron beads ($d = 549.8 \pm 6.8$ nm and $\zeta_p = -50.4 \pm 0.2$ mV at 0.1 mM LiCl) when low and high salt concentrations are $c_L = 0.1$ mM and $c_L = 10$ mM, respectively. A reference coordinate system is introduced as showed in (Figure 1A), with the origin located at the downstream junction and at the mid-point along the channel depth and width.

In the absence of a salt concentration gradient ($\nabla c = 0$) – namely, when all streams have the same low salinity level – the particles remain homogeneously dispersed in the colloidal bulk (Figure 1B). However, when the outer flow streams are switched to a high salt solution, a steady state salt gradient is imposed ($\nabla c \neq 0$), driving the particles towards higher salt concentration regions along this direction, as illustrated by the dashed lines in Figure 1A. This motion is driven by the $x$ component of the diffusiophoretic velocity, $u_{\mathrm{DP},x} = \Gamma_{\mathrm{DP}}\frac{\partial \ln c}{\partial x}$, with $\Gamma_{\mathrm{DP}}$ the diffusiophoresis mobility. In addition, two highly intense peaks of accumulated colloids, separated by a distance $\Delta_{\mathrm{p}}$, form at the entrance of the junction ($z/w_m = 0$), and rapidly converge towards the center of the device, downstream of the junction. The dynamics of these accumulation peaks have been studied in our previous works (18, 27), and Migacz and co-workers (35) have clarified further the underlying mechanisms. Briefly, the vertical ($y$-axis) component of the salt concentration gradient induces the formation of two symmetric accumulation peaks in close proximity to the top and the bottom walls of the microchannel. The dynamics of these peaks along the transverse direction is governed by the competition between the horizontal ($x$-axis) particle diffusiophoresis velocity, $u_{\mathrm{DP},x}$, which pushes the particle outwards, and the wall diffusioosmosis velocity, $u_{\mathrm{DO},x} = -\Gamma_{\mathrm{DO}}\frac{\partial \ln c}{\partial x}$ – with $\Gamma_{\mathrm{DO}}$ the diffusiophoresis mobility – which advects the particle inwards. The total transversal particle velocity can be approximated as $u_{p,x} = \Gamma_{\mathrm{eff}}\frac{\partial \ln c}{\partial x}$, where $\Gamma_{\mathrm{eff}} = \Gamma_{\mathrm{DP}} - \Gamma_{\mathrm{DO}}$ is the effective mobility. Hence, when $\Gamma_{\mathrm{eff}} > 0$, diffusiophoresis is dominant and the particles accumulated near the wall migrate towards the outer regions of the channel. Conversely, when $\Gamma_{\mathrm{eff}} < 0$, diffusioosmosis is dominant and the particles migrate towards the central region of the channel. For the particle type shown in Figure 1B (i.e., 0.5 $\mu$m, red), the effective mobility $\Gamma_{\mathrm{eff}}$ is negative for any salt concentration in the range from $c_L = 0.1$ mM to $c_H = 10$ mM, as shown in Figure S4. Consequently, this causes the displacement of the particle accumulation peaks towards the center, as it can be observed in the fluorescence micrographs in Figure 1B.

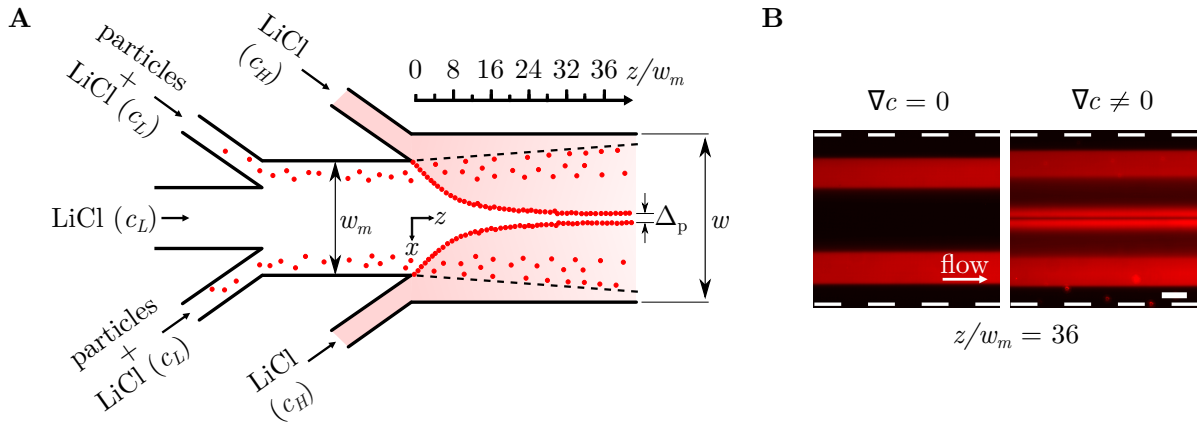

**Fig. 1. The double Ψ-junction microfluidic device.** (**A**) Schematic of the micro-chip for the manipulation of colloids via a salinity gradient. Inner inlet stream: LiCl solution at a low concentration ($c_L = 1\,\text{mM}$). Outer streams of upstream junction: LiCl solution at low concentration ($c_L$) seeded with red fluorescent ($d = 0.5\,\mu\text{m}$) carboxylate polystyrene particles, shown as red dots. Outer streams of downstream junction: LiCl solution at high salt concentration ($c_H = 1\,\text{mM}$). Dashed lines represent the outer boundaries of the colloidal streams. The red shading represents the local level of salinity. (**B**) Epi-fluorescence micrographs at $z/w_m = 36$ without (left) and with (right) a salinity gradient. Scale bar is 75 $\mu$m. White dashed lines represent the boundary of the micro-channel.

**Effects of Ionic Strength on Particle Dynamics.** The ionic strength of the solution can alter the zeta potentials of both particles and channel walls, which in turn affect the diffusiophoretic and diffusioosmotic mobilities, respectively. Moreover, the ionic strength can drastically influence the thickness of the electric double layer and, consequently, the diffusiophoretic mobility. The investigation of the effect of ionic strength on the particle focusing mechanism was conducted by varying the intensity of the low ($c_L$) and high ($c_H$) salt concentration streams while maintaining a constant salt concentration ratio $c_H/c_L$, thereby an approximately constant relative solute gradient $\nabla(\ln c) = \nabla c/c$. This allows one to modify the effective mobility $\Gamma_{\text{eff}}$ without significantly altering $\nabla(\ln c)$. Thus, it is reasonable to expect that the formation and subsequent motion of particle peaks near the walls of the device could also be altered.

*Experimental Analysis.* The dynamics of carboxylate polystyrene particles of various nominal sizes (20 nm, 100 nm, 200 nm, $0.5\,\mu$m and $1\,\mu$m) was investigated. The low salt concentration of the streams injected in the upstream junction was varied to either $c_L$ = 0.01, 0.1 or 1 mM, while the salt concentration ratio was fixed at $c_H/c_L = 100$ by adjusting the higher salt concentration $c_H$ of the outer flows of the downstream junction. Figure 2A-C present the normalized fluorescence intensity profiles along the transverse direction $x$ at a normalised distance $z/w_m = 36$ (i.e., $z = 9\,\text{mm}$) downstream of the junction for various particle sizes under different values of $c_L$. The normalized fluorescence intensity profiles $I(x^*)$ with $x^* = x/w_m$, provide a measure of the particle concentration along $x$, averaged over the depth of field of the microscope system and normalized with respect to the concentration of the solution injected into the device. Specifically, $I_{\text{norm}}(x) \simeq 0$ where no particles are detected, whereas $I_{\text{norm}}(x) \simeq 1$ where the depth-average particle concentration is close to the concentration $n_0$ of the colloidal solution injected into the device (see Supplementary Text for further details). In the case of low ionic strength – namely, $c_L$ = 0.01 mM and $c_H$ = 1 mM, compactly expressed as 0.01–1 mM in the remainder of the paper – sub-micron particles do not form accumulation peaks, and they display only a weak particle drift towards the inner region of the channel (Figure 2A). A particle accumulation effect is apparent instead for $1\,\mu$m colloids for which well-defined peaks – with a normalized average peak intensity of 1.3 – migrate slightly inward into the channel. Irrespective of the size, no particles are detected in the region $|x|/w_m < 0.1$, indicating a rather weak mobility of the particles along the transverse direction. Conversely, under moderate ionic strength (0.1-10 mM, Figure 2B), the accumulation effect is substantially amplified for all investigated particle sizes. In the case of sub-micron particles, the intensity of the colloidal front becomes more apparent, and the intensity profiles display intense fluorescent peaks that are otherwise absent under low electrolyte concentrations. Intriguingly, for $1\,\mu$m colloids, the normalized average peak intensity reaches a value of 2.9, almost two times greater than the peak intensity for the same particle size under low ionic strength condition. Consistently with the findings reported in our previous study (18), particles with similar zeta potential but larger size exhibit more intense accumulation effects due to their higher diffusiophoretic mobilities, which promote faster migration toward the walls. This is because smaller particles have larger values of the electric double layer thickness to particle radius ratios $(1/(\kappa a))$, which hinder their diffusiophoretic migration. The diminished particle accumulation effect under lower ionic strength conditions could be attributed to a reduced diffusiophoresis mobility, which is responsible for the particle migration along the channel depth-wise ($y$) direction and the accumulation at the top and bottom walls of the device. This interpretation is supported by the numerical simulations reported in the Supplementary Text, and it justifies why the observed particle manipulation mechanism is severely restricted in the salt concentration range 0.01-1 mM. Therefore, such conditions are not suitable for particle manipulation applications.

The transversal motion of accumulated particles – either in-ward or outward – can be predicted by computing the effective mobility $\Gamma_{\text{eff}}$ as function of the salt concentration (Figure S4). For salt concentrations equal or lower than 1 mM, $\Gamma_{\text{eff}}$ is always negative for any particles size. This indicates that diffusioosmotic induced flows prevail over particle diffusiophoresis in the $x$-direction. As a result, it can be anticipated that under low ionic strength conditions all particles will migrate inwards, as confirmed by the experimental observations in Figure 2A. Under moderate ionic strength conditions, the salt concentration varies between 0.1 to 10 mM, and in this range $\Gamma_{\text{eff}}$ is always negative for sub-micron particles. Again, the predicted inward migration of particles is confirmed by the experimental observations in Figure 2B. Remarkably, in the salt concentration range for moderate ionic strength (0.1-10 mM), the calculated effective mobility of $1\,\mu$m particles is negative at lower concentrations and positive at higher concentrations. Consequently, it is expected that the particles which are accumulated near the walls in the lower concentration region, migrate inward. Conversely, the particles which are accumulated near the walls in the higher concentration regions, migrate outward. This interpretation, which is also confirmed by the numerical model discussed in the next section, is consistent with the experimental data in Figure 2B. Indeed, the fluorescence intensity profile for $1\,\mu$m particles under moderate ionic strength conditions is the only profile displaying four accumulation peaks, two in the inner (lower salt concentration) region and two in the outer (higher salt concentration) regions. Importantly, the formation of four peaks moving along opposite directions cannot occur if the particle and wall zeta potentials and thus the diffusioosmosis, diffusiophoresis and effective mobilities were independent of the local salt concentration, as assumed in previous studies. Such finding demonstrates the importance of accounting for the effect of salt concentration on particle and wall zeta potentials and mobilities to interpret the experimental observations in our system.

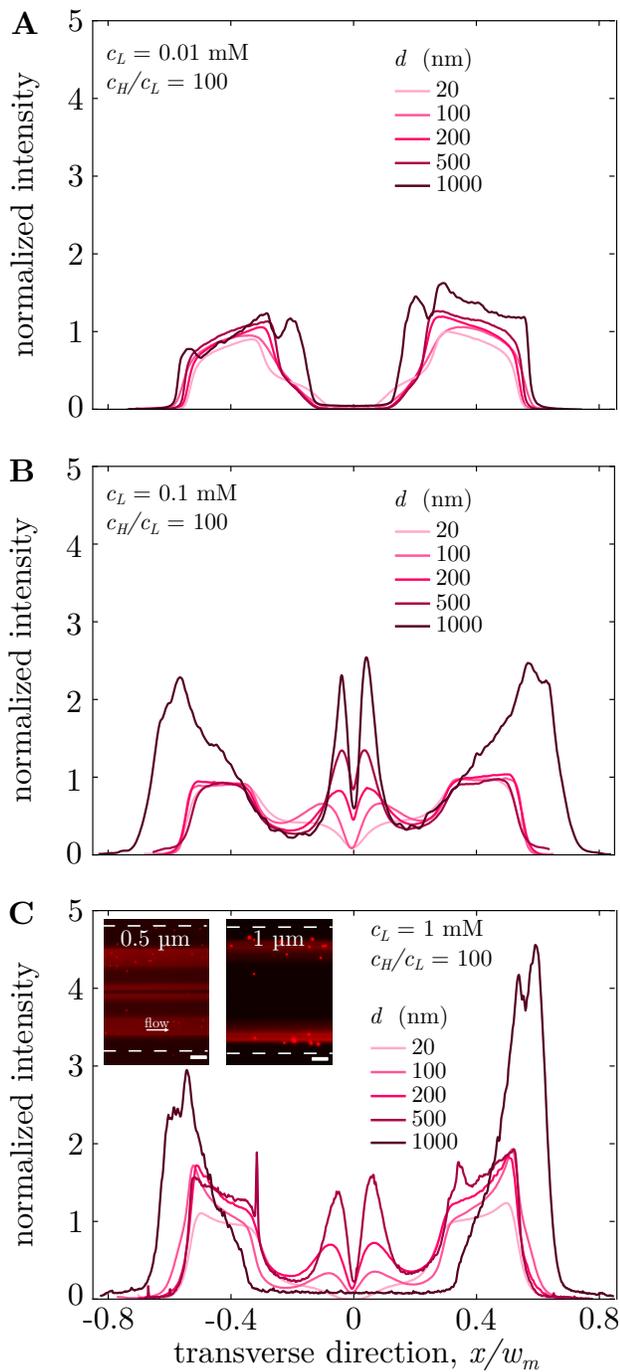

**Fig. 2. Normalized transverse fluorescent intensity profiles for particles of varying sizes under different ionic strength conditions.** Each panel corresponds to different ion concentration fields where $c_L$ is (**A**) 0.01 mM, (**B**) 0.1 mM, and (**C**) 1 mM, with a fixed concentration ratio $\frac{c_H}{c_L} = 100$. Fluorescent intensity profiles were analyzed from epi-fluorescent micrographs taken at a distance of $z/w_m = 36$ (i.e., $z = 9$ mm) downstream of the junction. Insets: fluorescence micrographs showing the different dynamics of $0.5\,\mu$m and $1\,\mu$m particles.

Figure 2C illustrates the intensity profile plots for the investigated particle size range under high ionic strength conditions (1-100 mM). For sub-micron particles, the profiles display four accumulations peaks – two in the inner region of the channel and two in the outer regions at the edges of the colloidal bands – with the exception of the 20 nm particles, for which the two inner peaks are not formed and a weak particle drift toward the center of the channel is observed instead. This observation is consistent with the calculated values of effective mobilities of these particles (Figure S4), which go from negative to positive at increasing salt concentrations within the range 1-100 mM. As observed for the moderate ionic strength case, the intensity of either diverging or converging peaks increases with the particle size because of the higher diffusiophoresis mobility of the larger particles. Moreover, the intensity of the peaks migrating inward is slightly lower under high ionic strength conditions than under moderate ionic strength conditions. Indeed, due to the change in the sign of the effective mobilities under high ionic strength conditions, a fraction of the colloids accumulated at the walls is pushed outward, resulting in a lower amount of colloids focusing in the inner region of the channel.

For $1\,\mu$m colloids, the accumulation peaks are absent from the center of the device (Figure 2C). Instead, two intense peaks can be seen to diverge outward, towards high salt concentration regions (Figure 2C). This behavior is in stark contrast to the inward transverse displacement displayed by all sub-micron particles ($\leq 500$ nm) tested under the same ion concentration field. The observed particle dynamics aligns with our interpretation of a salt concentration-dependent focusing mechanism since for $1\,\mu$m particles the effective mobility is positive for salt concentrations above $3.4\,$mM, namely almost the whole range of concentration from 1 to 100 mM. Strikingly, the diverging particle peaks exhibit significantly pronounced fluorescent intensities with a normalized maximum peak intensity value of 4.55, which is much larger than the intensity peak values achieved for any other particle size under all examined conditions. Such an intense focusing effect is the result of the high values of the diffusiophoresis mobility of the $1\,\mu$m particles in the 1-100 mM salt concentration range (Figure S3). Specifically, in this concentration range the diffusiophoresis mobility of the $1\,\mu$m particles reaches a maximum of $960\,\mu$m$^2$s$^{-1}$, that is more than twice the mobility of any sub-micron particles at any examined salt concentration.

To conclude, it is apparent that a high electrolyte concentration field induces a change in the trajectory of $1\,\mu$m particle peaks when compared to smaller particle sizes (Figure 2C,D). The difference in particle dynamics implies that this effect can be exploited for separation applications, as exemplified in the micrographs in Figure 2C illustrating the contrasting motion of 0.5 and $1\,\mu$m particle peaks under high ionic strength conditions. Before discussing the implementation of this particle separation strategy, we will present the results of the numerical study that was conducted to confirm the interpretation of the experimental analysis of particle dynamics under high ionic strength conditions.

***Numerical Analysis.*** The dynamics of the colloidal particles was simulated through a finite element model developed in COMSOL Multiphysics (Supplementary Text). The numerical hydrodynamic velocity field $\boldsymbol{u}$, salt concentration $c$, and particle concentration $n$, were calculated in a 3D domain consisting of a straight rectangular channel. For the domain shape, we adopted the channel sizes of the revised double-junction device, used in the separation experiments described in the next sections. In this new device, the average velocity is the same for all inner and side channels, thus there is no hydrodynamic broadening of the streams, which occurs instead in the version of the device used for the experiments in Figure 2. To simulate the dynamics of the different types of particles examined in the experiments, the particle diffusiophoresis mobilities were calculated as a function of the local salt concentration as detailed in the Supplementary Text. Similarly, a salt concentration-dependent diffusioosmosis mobility was used in the simulation as detailed in the Supplementary Text.

Figure 3B shows the simulated transverse profile of the normalized particle concentration $n/n_0$ for $0.5\,\mu$m red fluorescent colloids at a normalized distance from the downstream junction of $z/w_m = 25$ (i.e., $z = 5$ mm) under high ionic strength conditions (1-100 mM). Note that the numerical profiles are calculated from the simulated particle concentration field by averaging the concentration over the channel depth ($y$ direction). Consistently with the experimental observations in Figure 2C, the simulated profile displays four accumulations peaks, two in the inner region and two in the outer regions of the channel. A very good quantitative agreement between numerical predictions and experimental observations can be seen in the inset of Figure 3B for the normalized inward displacement of the inner peaks, defined as $\delta_p(z) = (\Delta_p(0) - \Delta_p(z))/2\,w_m$. Figure 3C shows the simulated particle concentration field on the plane perpendicular to the flow direction at $z/w_m = 25$. Due to symmetry with respect to the $x$ and $y$ axes, the concentration field is visualized only on the top right quadrant of the channel cross-section, corresponding to the red-shaded region in the schematic in Figure 3A. The competing effects of diffusioosmosis and diffusiophoresis at the channel walls are apparent in the in-plane total particle velocity field $\boldsymbol{u}_\text{p}$, shown by the white arrows in Figure 3C. As expected, since the particle effective mobility is negative (positive) in the low (high) salt concentration regions, the total particle velocity near the wall is directed inward (outward) in the inner (outer) region of the channel. Consequently, a fraction of the particles accumulated near the wall migrate inward and the reminder outward, leading to the formation of two converging and two diverging peaks in the simulated particle concentration profile (Figure 3B), in agreement with the experimental observations (Figure 2C).

On the other hand, for $1\,\mu$m particles under high ionic strength conditions the effective mobility is positive, and the total particle velocity is directed outward, almost anywhere near the walls (Figure 3E). Consequently, all particles accumulated at the wall migrate outward, resulting in the formation of only two intense diverging peaks in the outer regions of the channel, which are visible both in the simulated particle concentration profile in Figure 3D and in the experimental fluorescence intensity profile in Figure 2C. Furthermore, in the simulations – as observed in the experiments – the $1\,\mu$m colloids are totally excluded from the central region of the channel $|x|/w_m < 0.25$. Notably, the high value of diffusiophoresis mobility of $1\,\mu$m particles at the examined salt concentrations leads to a strong accumulation of colloids near the wall, where the particle concentration is predicted to exceed 30 times that of the solution injected into the device (Figure 3E).

**Separation of Particles by Surface Chemistry.** The capability of continuously separating $0.5\,\mu$m from $1\,\mu$m carboxylate polystyrene particles by diffusiophoresis is remarkable, but also quite surprising. Indeed, in these particle size range the diffusiophoresis mobility weakly depends on particle diameter, which suggests that the separation mechanism must be chemical in nature. On the other hand, the particles are made of the same polymer modified with the same functional group, hence no significant difference in the particle chemistry is expected. In fact, at a salt concentration of $c_L = 1\,$mM, the two particle species have a similar zeta potential, namely $\zeta_p = -47.2 \pm 0.4$ mV for $0.5\,\mu$m and $\zeta_p = -68.7 \pm 1.1\,$mV for $1\,\mu$m colloids, corresponding to a relative difference of about 30%.

The solution to this conundrum lies in the plots of particle zeta potential against salt concentration (Figure 4), which are noticeably different for these two types of particles. Specif-ically, for $0.5\,\mu$m red particles the magnitude of the zeta potential decreases monotonically with the logarithm of the electrolyte concentration (i.e. $\partial|\zeta|/\partial c < 0$), whereas for $1\,\mu$m particles the zeta potential has a point of maximum and below this point its magnitude increases with the salt concentration (i.e. $\partial|\zeta|/\partial c > 0$). The fall off in the magnitude of zeta potential of polystyrene colloids at low concentrations of an indifferent monovalent electrolyte has been investigated in several studies (36–43). This observation contradicts the predictions of the classical Gouy–Chapman model of the electrical double layer for which a monotonic decrease with increasing salt concentration is expected due to the compression of the electric double layer. A similar trend has been observed also for other materials, such as silica (44) and other minerals (45). It has been shown that the weakening of the zeta potential, measured by electrophoretic light scattering, at the lower ionic strengths reflects the reduction in electrophoretic mobility caused by the surface electrical conductivity of the polystyrene particles. This effect becomes apparent at the lower salt concentrations (i.e. lower conductivities of the surrounding medium $\sigma_m$) when the surface conductance of the particles $K_s$ is comparable to the product of the medium conductivity and particle radius $a$ - namely, the Dukhin number $K_s/\sigma_m a$ is not negligible. Combined with the compression of the electric double layer at the higher ionic strengths, the electrophoretic mobility and, thus the zeta potential calculated according to the classical Gouy-Chapman model, may exhibit a point of maximum. The effects of surface conductance on particle electrophoretic mobility are more prominent for larger particles (41) and higher surface charges (37). This justifies why the $1\,\mu$m particles with twice the size and four times the surface concentration of carboxyl groups compared to the red $0.5\,\mu$m particles (Table S1), display a point of maximum in the zeta potential plot. Conversely, for all other red sub-micron particles, the effect of surface conductance is negligible and the zeta potential is instead a monotonic func-tion of the salt concentration. In the next sections, we demonstrate the effective separation of similarly-sized carboxylate polystyrene particles based on their opposite zeta potential sensitivity to the local salt concentration.

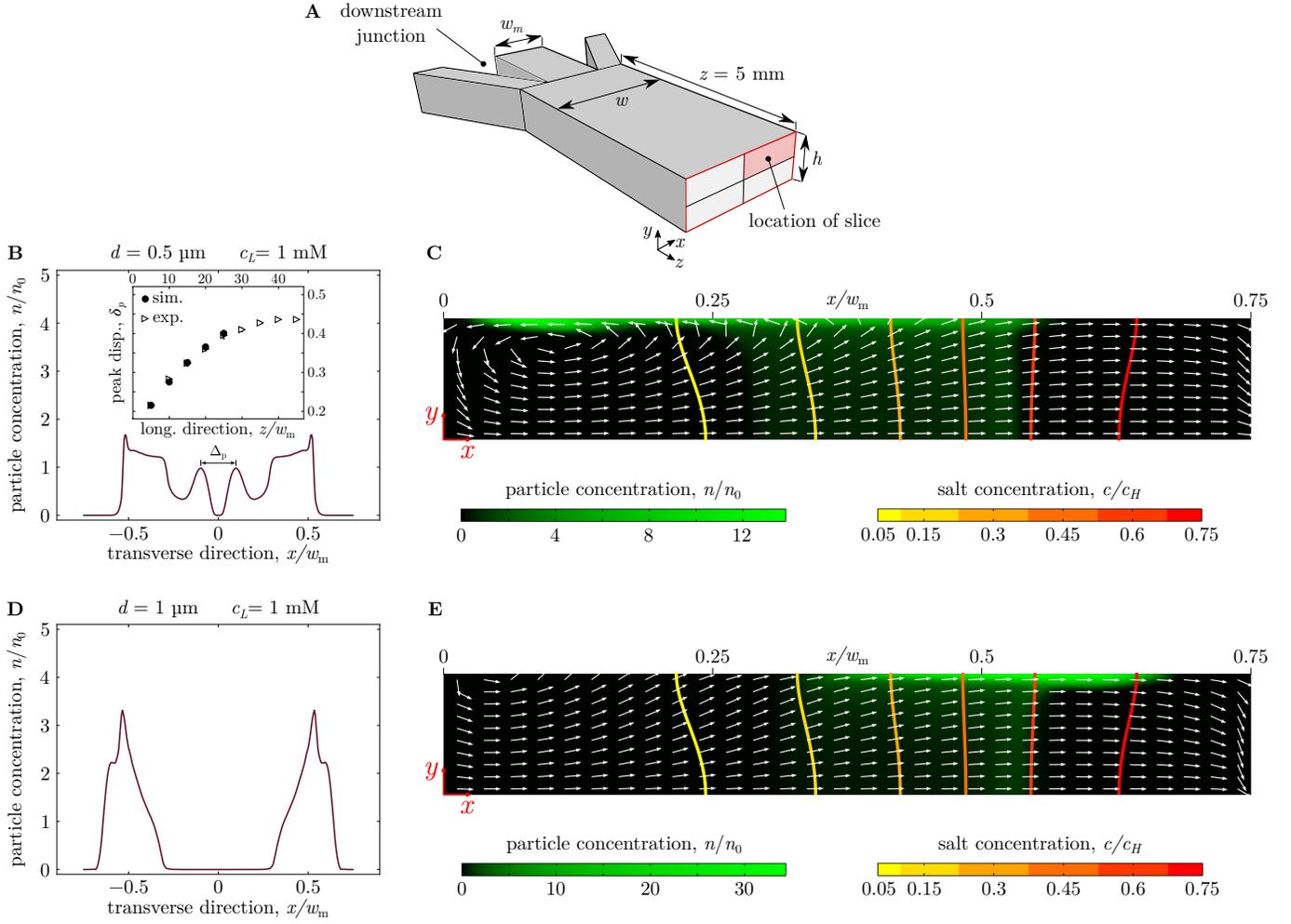

**Fig. 3. Numerical simulation results.** (**A**) Schematic of the channel in the region near the downstream junction where particle dynamics were simulated ($w_m = 300\,\mu$m, $h = 45\,\mu$m). (**B,C**) Simulated normalized particle concentration profile (**B**) and particle distribution (**C**) on the cross-section slice highlighted in red in panel (**A**) at $z/w_m = 25$ (i.e., $z = 5$ mm) for $0.5\,\mu$m red particles at high ionic strength condition (1-100 mM). Inset in panel (**B**) shows the simulated (solid circles) and experimental (open triangle) normalized inward displacements of the inner peaks, $\delta_p = (\Delta_p(0) - \Delta_p(z))/2\,w_m$, as a function of the normalized distance from the downstream junction, $z/w_m$. (**D,E**) Simulated normalized particle concentration profile (**D**) and particle distribution (**E**) on the same cross-section slice at $z/w_m = 25$ (i.e., $z = 5$ mm) for $1\,\mu$m red particles at high ionic strength condition. In panel (**C**) and (**E**), the white arrows represent the streamlines of the total particle velocity $\boldsymbol{u}_p$ and the solid colored lines are the salt concentration isolines, which are slightly bent due to the Poiseuille-like velocity profile, thus leading to the generation of a vertical component of the salt concentration gradient.

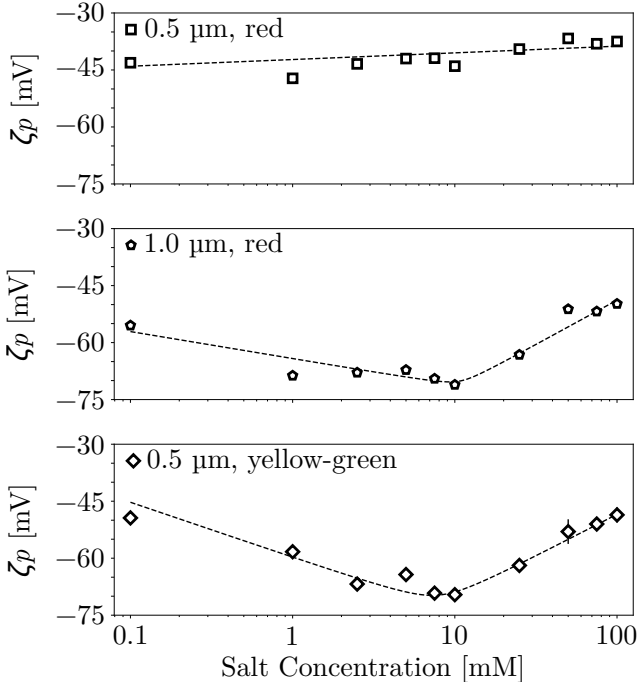

**Fig. 4. Zeta potential $\zeta_P$ as a function of LiCl salt concentration for different particle types**. Top: $0.5\,\mu m$ fluorescent red. Middle: $1\,\mu m$ fluorescent red. Bottom: $0.5\,\mu m$ fluorescent yellow-green particles. Markers are experimental data and dashed solid lines are the fits to Eq. (S3) in Supplementary Text for the $0.5\,\mu m$ red particles and to Eq. (S4) in Supplementary Text for the $1\,\mu m$ red and $0.5\,\mu m$ yellow-green particles. Error bars are calculated as the standard deviation of three repeated measurements of a sample.

***Separation of Red $0.5\,\mu m$ and $1\,\mu m$ Particles.*** For the separation experiments, we introduced a new variation of the double $\Psi$-junction device that enables the separation and extraction of particles converging towards the center of the device. Figure 5A illustrates a schematic of the updated device with two key modifications. First, two side outlets and one central outlet are incorporated at the end of the microfluidic chip. The central outlet is designed to be quite narrow to prevent the extraction of undesired streams. The design of the outlet channels ensures that under a no salinity gradient condition ($\nabla c = 0$), the injected colloidal streams exit solely through the side channel outlets. On the other hand, when there is a salinity gradient ($\nabla c \neq 0$), the particles from the peaks that converge towards the inner region of the device, can be extracted and recovered from the central outlet stream. Furthermore, the length of the channel downstream of the second junction was increased to allow the separation process to occur over longer distances. Finally, the width of the inner and outer channels of the two junctions were adjusted to prevent hydrodynamic focusing or broadening (18, 46) (see Supplementary Text for channel sizes).

To carry out the separation experiment, a salt concentration ratio of $c_H/c_L = 100$ was imposed under high ionic strength conditions (1-100 mM). The flow configuration remained identical to previous experiments, as depicted in Figure 5A, but the colloidal solution injected into the side channels of the downstream junction consisted in a binary mixture of $0.5\,\mu m$ diameter (0.02% v/v) and $1\,\mu m$ diameter (0.002% v/v) red fluorescent carboxylate polystyrene particles. The flow rate in each inlet of the device was $10\,\mu l/min$. To evaluate the separation process, the colloidal mixture injected into the device, referred to as the inlet stream, and the solution extracted from central outlet were analyzed. Note that since both particle species were stained with the same fluorophore dye, it was not possible to perform an on-chip analysis of the separation process via epi-fluorescence imaging. Therefore, an off-line epi-fluorescence imaging procedure was adopted instead (see *Materials and Methods* section for details). Figure 5B shows an exemplar off-chip epi-fluorescence image of the dried colloidal sample extracted from the inlet (left) and central outlet (right) solutions. Note that the strong refractive index contrast between the colloidal particles and the surrounding medium (air) resulted in image distortion and particle lens effect, explaining why the $1\,\mu m$ particles appear much larger than the $0.5\,\mu m$ particles. Analyzing the inlet mixture sample revealed a total particle count of 11,629 where 90.5% of the population is represented by $0.5\,\mu m$ colloids and the remaining 9.5% being $1\,\mu m$ particles. Conversely, a total of 4,221 particles were detected in the central outlet droplet, all of which were $0.5\,\mu m$ colloids. This corresponds to a 100% removal of $1\,\mu m$ particle from the central outlet stream and a 40% recovery rate of $0.5\,\mu m$ particle, namely 40% of the smaller particles injected into the device are recovered in the central outlet stream. The successful separation of the smaller particles from the mixture was also validated via dynamic light scattering analysis of the inlet and central outlet solutions (see Supplementary Text).

***Separation of $0.5\,\mu m$ Red and Yellow-green $0.5\,\mu m$ Particles.*** In the previous experiment, the particle separation is achieved because the opposite sensitivities of particle zeta potential to salt concentration results in opposite effective mobilities (i.e. $\Gamma_{\text{eff}} < 0$ for sub-micron particles and $\Gamma_{\text{eff}} > 0$ for $1\,\mu m$ particles) at almost any concentration in the range from $c_L$ to $c_H$ under higher ionic strength conditions (1-100 mM). The separation is not possible at lower and intermediate ionic strength conditions (i.e., 0.01-1 mM and 0.1-10 mM), since the effective mobility of $1\,\mu m$ particles is also mostly negative at these salt concentrations. Nevertheless, it can be hypothesized that two particle species can still be separated even if they exhibit a negative effective mobility in the lowest salt concentration stream ($c_L$), namely they have rather similar diffusiophoresis mobilities, and hence similar size and zeta potential in this stream. Indeed, as the electrolyte streams with low ($c_L$) and high ($c_H$) salt concentrations mix downstream of the junction, the particles are exposed to a surrounding solution with increasing salinity levels. If the rise in local salt concentration affects the zeta potential of the two particle types in different ways, then their effective mobilities may also evolve differently depending on the particle surface physico-chemical properties. Specifically, if the local salt concentration becomes high enough to reverse the sign of the effective mobility of one particle species but not the other, the particles can be separated. Such a strategy would enable the fine separation of particles with very similar size and zeta potential at the lowest salt concentration $c_L$, but different surface physico-chemical properties that affect their zeta potential sensitivity to local ionic strength.

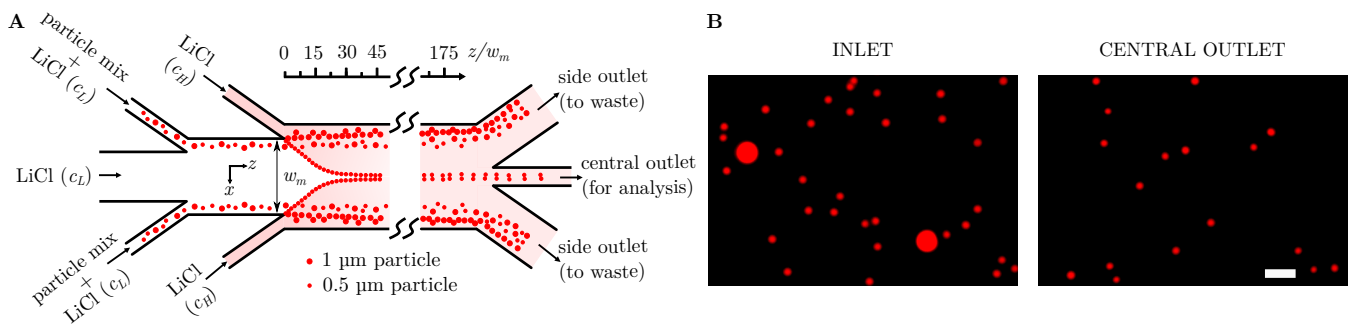

**Fig. 5. Separation of a binary suspension of red fluorescent 0.5 $\mu$m and 1 $\mu$m carboxylate polystyrene particles.** (**A**) Schematic of a long double $\Psi$-junction device showcasing the separation and subsequent extraction of 0.5 $\mu$m from 1 $\mu$m particles under high ionic strength conditions. A low salt concentration stream of $c_L$ = 1 mM was injected in all channels of the upstream junction with the side channels containing a binary 1:10 (v/v) mixture of 1 $\mu$m and 0.5 $\mu$m carboxylate polystyrene particles. High salt concentration ($c_H$ = 100 mM) streams were introduced in the side channels of the downstream junction. (**B**) Off-chip epi-fluorescence images of dried samples of the inlet (left) and central outlet (right) streams. Scale bar represents 2.5 $\mu$m. The channel width $w_m$ is 200 $\mu$m.

To demonstrate this separation strategy, an experiment was conducted in a double $\Psi$-junction device with the revised geometry previously described. The flow configuration was as in Figure 5a with $c_L = 1$ mM and $c_H/c_L = 100$. In this instance, two populations of $0.5\,\mu$m carboxylate polystyrene particles, one fluorescently stained in red and one in yellow-green, were mixed in 1:1 (v/v) ratio in a solution at 0.002% v/v concentration and injected in the side channels of the upstream junction. The flow rate in each inlet of the device was 3.65 $\mu$l/min. At a salt concentration of $c_L = 1$ mM, the $0.5\,\mu$m red and yellow-green particles have similar zeta potential, namely $\zeta_p = -47.2 \pm 0.4$ mV for red and $\zeta_p = -58.3 \pm 1.1$ mV for yellow-green colloids, corresponding to a relative difference of just 20%. As a result, the effective mobility is negative for both particle types at this salt concentration. Importantly, the two particle species, although made of the same material, have a different surface chemical composition whereby the surface concentration of carboxyl groups of the yellow-green particles is almost nine times the one of the red particles (Table S1). Consequently, the effects of particle surface conductance are not negligible for the yellow-green particles and the magnitude of the zeta potential, and thus of the electrophoretic mobility, increases with the salt concentration until it reaches a maximum around 10 mM (Figure 4).

Figure 6 shows the individual and combined particle dynamics of both sets of colloids in the mixture under high ionic strength conditions. Epi-fluorescence micrographs of either red (Figure 6A) or yellow-green (Figure 6B) particles were taken at short (i.e. $z/w_m = 45$, namely $z = 9$ mm) and long (i.e. $z/w_m = 175$, namely $z = 35$ mm) distances from the downstream junction. At a short distance, the normalized fluorescence intensity profiles along the transverse direction $x$ (Figure 6C) exhibit a similar trend to those in Figure 2C for sub-micron colloids under high ionic strength conditions at a distance of $z = 9$ mm from the downstream junction. Since for both particle types the effective mobility change signs within the salt concentration range 1-100 mM, two converging and two diverging accumulation peaks can be observed in the intensity profiles for both red and yellow-green particles. The slight difference in the intensities and positions of the four peaks between the two profiles is due to the different diffusiophoresis mobilities of the two particle species. At such a distance from the junction, the regions of the channel occupied by the red and yellow-green particles overlap and separation is not possible. On the other hand, a stark difference between the intensity profiles of the two particle types is apparent at a long distance from the junction. At this location, the two peaks of yellow-green colloids, originally located in the inner region of the channel, almost vanished. Conversely, the two red particle peaks in the centre of the channel survived, retaining a similar intensity to the one observed closer to the junction. Consequently, the near-complete disappearance of focused yellow-green particles in the inner region of the channel enables the separation between the two particle species. As illustrated by the overlaid red and yellow-green micrographs in Figure 6D, the red particle can be extracted from the binary mixture of red and yellow-green particles and recovered from central outlet stream.

To justify the observed dynamics, a numerical simulation was performed to calculate the salt concentration profile along the transverse direction at the channel wall and at either short ($z/w_m = 45$) or long ($z/w_m = 175$) distances from the downstream junction (Supplementary Text). The profiles were then used to calculate the effective mobility as a function of the transverse direction $x$ for both red and yellow-green particles. As shown in Figure S9a-b, the effective mobility of the red particles is negative in the inner region of the channel $x/w_m < 0.25$ both at short and long distance from the junction. On the other hand, the effective mobility of yellow-green particles in the same region changes from negative values at $z/w_m = 45$ to positive values at $z/w_m = 175$. This is because at large distances from the junction the electrolyte streams are mixed and the salt concentration at the channel walls ranges approximately between 2 and 3.5 mM. At these concentrations, the magnitude of the zeta potential of the yellow-green particles increases ($\partial|\zeta|/\partial c > 0$) and their diffusiophoresis mobility is high enough to overcome the diffusioosmosis mobility, causing the particles to leave the center of the channel and migrate outwards. This is not the case for the red particles, for which $\partial|\zeta|/\partial c < 0$ and the diffusiophoresis mobility depends very weakly on the salt concentration (Figure S3). As a result, the dominating effect of diffusioosmosis confines the red particles in the inner region of the channel.

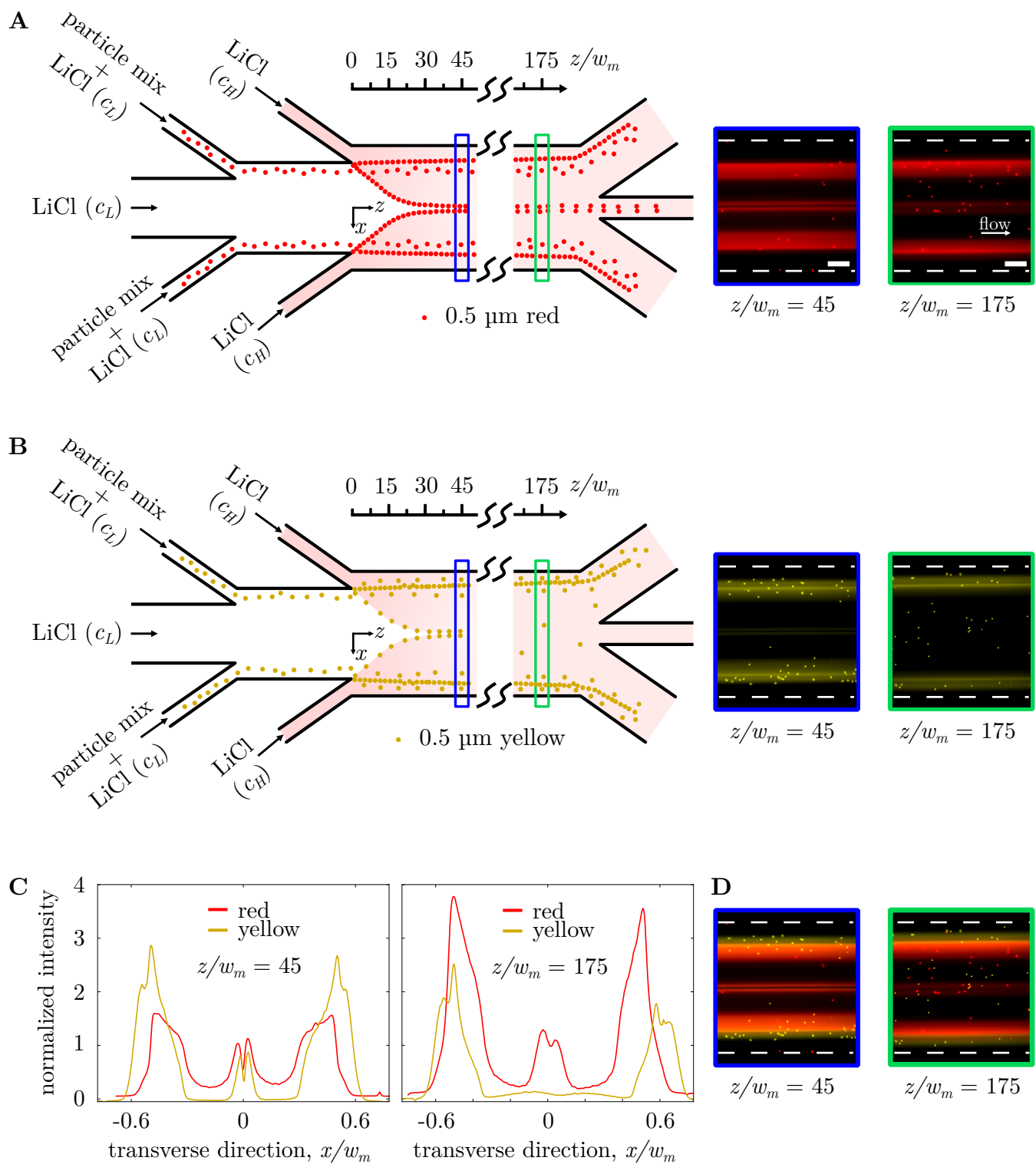

**Fig. 6. Separation of a binary 1:1 (v/v) mixture of red-fluorescent and yellow-green fluorescent 0.5 μm carboxylate polystyrene particles.** A low salt concentration ($c_L = 1\,\text{mM}$) was injected in all channels of the upstream junction with the side channels containing the binary colloid mixture. High salt concentration ($c_H = 100\,\text{mM}$) streams were injected in the side channels of the downstream junction. (**A**,**B**) Schematic and epi-fluorescence micrographs illustrating the distribution of red (**A**) and yellow-green (**B**) 0.5 μm particles at short ($z/w_m = 45$) distance (blue rectangle) and long ($z/w_m = 175$) distance (green rectangle) from the downstream junction. (**C**) Normalized fluorescence intensity profiles for both red and yellow-green particles at $z/w_m = 45$ (left) and $z/w_m = 175$ (right). (**D**) Overlaid epi-fluorescence micrographs of red and yellow-green 0.5 μm particles at $z/w_m = 45$ (left) and 175 (right). Scale bar represents 50 μm.

## Discussion

Our findings demonstrate that the effect of salt concentration on the zeta potentials of the channel walls and colloidal particles can affect significantly the particle dynamics in our microfluidic device under the examined experimental conditions. A constant zeta potential model ($\partial\zeta/\partial c$=0), which is commonly used in diffusiophoresis studies, cannot explain characteristic features of the observed particle dynamics, such as the formation of four accumulation peaks in the particle concentration profiles along the transverse direction of the channel (Figure 2) and the reversal of the direction of motion of the accumulation peaks downstream of the junction (Figure 6). At the low ionic strength conditions (0.01-1 mM), particle manipulation capability is severely restricted due to the weakening of diffusiophoresis effect caused by the thickening of the electric double layer. This result aligns with prior research, wherein, for a constant salt concentration ratio, lower electrolyte concentrations led to lower diffusiophoresis mobilities (47, 48).

Generally, the accumulation of particles at the channel walls is more intense for larger particles. For $1\,\mu$m particles under high ionic strength conditions, we predicted a peak of particle concentration 30 times larger than the concentration of the colloidal solution injected into the device – an effect that could be exploited for the preconcentration of samples in microfluidic analytical systems. Particle focusing is only weakly affected by the particle size since the diffusiophoresis mobility depends weakly on this parameter. Under high ionic strength conditions, the average magnitude of the accumulation peaks in the transverse intensity profile for the $500\,$nm particles (ca. 1.5) is about four time higher than the peak's magnitude for the $100\,$nm particles (ca. 0.35) (Figure 2C). This result suggest that the device could be adopted for the continuous removal of $0.1\,\mu$m particles from a mixture of $0.1\,\mu$m and $0.5\,\mu$m colloids, but the separation process would be incomplete since a fraction of the smaller particles will remain mixed with the larger particles.

Remarkably, the diffusiophoresis-driven dynamics of the particles is significantly affected by the surface concentration of carboxyl groups even if the particles have similar sizes and similar values of zeta potential in the solution in which they are originally dispersed (i.e. the low salt concentration stream). At first, this may sound like an oxymoron since zeta potential and diameter are the only properties of a rigid particle that affect its diffusiophoresis mobility. However, for larger ($\geq 0.5\,\mu$m) particles, a higher surface concentration of ionizable groups translates, at the lower ionic strengths, in a non-negligible surface conductance effect that slows down the electrophoretic motion of the particle (40). Consequently, for particles with a higher surface concentration of carboxyl groups, the magnitude of the zeta potential (measured via electrophoretic light scattering) increases with the local salt concentration ($\partial|\zeta|/\partial c > 0$) until it reaches a point of maximum between 1 and $10\,$mM. Conversely, the particles with a lower surface concentration of carboxyl groups exhibit a magnitude of the zeta potential that decreases monotonically with the salt concentration ($\partial|\zeta|/\partial c < 0$). As the electrolyte streams mix downstream of the junction, the particle zeta potential and, thus, diffusiophoresis mobility evolve differently depending on the particle's carboxyl group surface concentration.

Importantly, in our device the direction of particle migration along the transverse direction depends on the balance between the convection flows induced by diffusioosmosis at the channel walls, pushing the particle inwards, and the particle diffusiophoresis motion, pushing the particle outwards. Therefore, although the diffusiophoresis mobility of the examined particles is always positive, as they are all negatively charged, the colloids can move in opposite directions (either inwards or outwards) and, thus they can be separated, depending on whether their local diffusiophoresis mobility is greater or lower than the local diffusioosmosis mobility. In other words, diffusioosmosis at the channel walls plays a critical role in the separation mechanism. In previous studies on passive and continuous separation of particles by diffusiophoresis the role of diffusioosmosis was negligible, and only particles with oppositely charged surface groups, and thus opposite diffusiophoresis mobilities, could be separated easily. In an externally-activated system, like the one recently introduced by Bekir and co-workers (49), diffusioosmostic flows are instead instrumental to the particle manipulation mechanism and particles can be separated based on their porosity, bulk chemistry or type of surface groups; however, this approach applies only to sedimenting particles and, like other active manipulation technique, it requires the use of an external energy source and auxiliary equipment. In our device, polystyrene particles with similar or identical sizes and same surface group chemistry, exhibiting only a slight difference (up to 30%) in the zeta potential in the solution in which they are dispersed, can be continuously separated with 100% efficiency solely based on the concentration of the surface carboxyl groups. This result is unprecedented as no other microfluidic technique, whether active or passive, enables continuous particle separation based solely on the concentration of a surface group.

The proposed mechanism opens up new opportunities for the continuous separation of particles based not only on the surface concentration of ionizable groups but also on other important surface physico-chemical properties that influence the sensitivity of the particle zeta potential to the local electrolyte concentration. For instance, it is reported that the roughness of polystyrene microbeads can affect their electrokinetic properties and, thus, zeta potential depending on the ionic strength of the solution (38, 50). Hence, the mechanism reported in this study could be exploited for the separation of rough particles, for which $\partial|\zeta|/\partial c > 0$, from smooth particles, for which $\partial|\zeta|/\partial c < 0$ (42). Similarly, particles coated with a polyelectrolyte coating, for which $\partial|\zeta|/\partial c \simeq 0$, could be separated from uncoated particles for which $\partial|\zeta|/\partial c < 0$ (32). Furthermore, particle surface chemistry could be distinguished based on their different affinity to ions and other solutes that may affect their zeta potential to different extent under varying ionic strength conditions (42). More generally, the effects of 3D charge distribution, surface ion and flow permeabilities (i.e. porosity) and surface heterogeneity may alter the sensitivity of the particle zeta potential and, thus, of the electrophoretic mobility to the local salt concentration (51). The proposed microfluidic strategy could be then potentially used for the continuous separation of colloids, such as engineered particles, environmental contaminants and biocolloids (e.g., liposomes, extracellular vesicles and bacterial cells) distinguished solely by differences in these surface physical and chemical properties, which is of great relevance especially for the purification of synthetic and biological samples. By relying on a simple experimental set-up with no external energy sources and by adopting easy-to-operate microchips, our microfluidic technique may facilitate fundamental studies of the physics, chemistry and biology of these particles, and also underpin the development of new microfluidic technological tools for chemical and biochemical analysis, diagnostics, environmental monitoring, drug screening, and drug delivery applications.

## Materials and Methods

**Materials.** Polydimethylsiloxane (PDMS) RTV-615a and curing agent RTV-615b used for the fabrication of microfluidic channels were purchased from Techsil, UK. Pre-cleaned Corning™ glass slides used as the substrate to make PDMS microfluidic chips were purchased from ThermoFisher scientific. 1H,1H,2H,2H-Perfluorooctyl-trichlorosilane was purchased from Merck and used for substrate silanisation. All aqueous solutions were prepared using deionized water (18.2 M$\Omega$ cm) produced from an ultrapure milli-Q grade purification system (Millipore, USA). Lithium chloride salt (LiCl, 99%) was purchased from Acros Organics. Red fluorescent (580/605), carboxylate polystyrene beads (Invitrogen™ Part Number F8887, lot number 2294999) were purchased from ThermoFisher Scientific at various nominal sizes. The 200 nm red fluorescent (580/605), carboxylate polystyrene beads were purchased separately from ThermoFisher Scientific (Invitrogen™ Part Number F8810, lot number 1934417A). Also the 500 nm yellow-green fluorescent (505/515) carboxylate polystyrene beads were purchased separately from ThermoFisher Scientific (Invitrogen™ Part Number F8813, lot number 2247969).

**Sample Preparation.** For all microfluidic experiments, the colloidal particle stock solutions (2% solids) were diluted in either $c_L$ = 0.01 mM, 0.1 mM or 1 mM LiCl solution. The particle concentration was chosen to work in the dilute regime (i.e., negligible particle interaction) and to prevent both clustering and excessive surface contamination. The exact salt and particle concentrations of the dispersions used are defined in their respective Figure captions or in the main text.

**Fabrication and Operation of the Microfluidic Devices.** Microfluidic devices were made from PDMS channels bonded to a glass substrate. To make PDMS channels, standard photo-lithography and soft-lithography techniques were employed, as reported elsewhere (18). For the results reported in Figure 1 and 2, an inverted optical microscope (TE300, Nikon) equipped with a 10x objective lens (0.25 NA) was used to capture epi-fluorescence micrographs of the devices. A fluorescent lamp (CoolLED pE300) was used to excite the sample, which allowed the collection of fluorescent intensity data via a CCD camera (Ximea MQ013MG-ON). Collected data is in the form of 1264 x 1016 px, 16 bit TIFF images. For the results reported in other Figures, a homemade inverted optical microscope was used to capture epi-fluorescence micrographs of the device. The microscope was assembled with Thorlabs components and featured an X-Cite 200 fluorescent lamp (Excelitas, Canada) and a CCD camera (Ximea MQ013MG-ON). A Nikon Plan fluor 10x objective lens (0.30 NA) was used for online imaging of particle dynamics inside the microfluidic devices. For the experiment requiring offline epi-fluorescence imaging (Figure 5), an oil immersion 100x (1.25 NA) was used instead. All epi-fluorescence micrographs of the microfluidic devices were acquired by positioning the focal plane nearby the bottom (glass) wall and subsequently processed via ImageJ (contrast enhancement, LUT color change).

**Offline Epi-fluorescence Micrograph Analysis.** A 30 µL drop of either the inlet or the central outlet solution was independently placed on a clean microscope slide. The droplet was then sandwiched by a coverslip and allowed to dry overnight. A Parafilm® layer with a punched circular all of ca. 5 mm in diameter was used as a spacer between the two glass substrates. Both the coverslip and the microscope slide were silanized via physical vapor deposition to make them hydrophobic and prevent the spreading of the drop on the glass surfaces. Different regions of the dried sample were then imaged under epi-fluorescence mode with hundreds of micrographs captured for each sample. ImageJ software was used to count the total number of particles in each frame and to determine the particle size distribution within the inlet and outlet solution.

**Particle Characterization.** Size and zeta potential measurements were performed a minimum of three times at 25 °C using a ZetaSizer Ultra Red (Malvern Panalytical) equipped with a He-Ne laser (633 nm). For size measurements, non-invasive backscatter (173° angle) was used for all particle sizes. Error bars for particle size and zeta potential were calculated as standard deviations of at least three consecutive measurements of the sample loaded in the ZetaSizer. Details on the zeta potential measurements are provided in Supplementary Text.

# Data Availability

The data underlying this study are openly available on University College London repository doi.org/XXXXXX.

**ACKNOWLEDGEMENTS**

This research was supported by Engineering and Physical Sciences Research Council (EPSRC) grants EP/S013865/1 and EP/X01813X/1. For the purpose of open access, the authors have applied a Creative Commons Attribution (CC BY) licence to any Author Accepted Manuscript version arising.

# Bibliography

1. Shujing Lin, Zixian Yu, Di Chen, Zhigang Wang, Jianmin Miao, Qichao Li, Daoyuan Zhang, Jie Song, and Daxiang Cui. Progress in microfluidics-based exosome separation and detection technologies for diagnostic applications. *Small*, 16(9):1903916, 2020.
2. Dominik Buschmann, Veronika Mussack, and James Brian Byrd. Separation, characterization, and standardization of extracellular vesicles for drug delivery applications. *Advanced drug delivery reviews*, 174:348–368, 2021.
3. Seyed Mohammad Majedi and Hian Kee Lee. Recent advances in the separation and quantification of metallic nanoparticles and ions in the environment. *TrAC Trends in Analytical Chemistry*, 75:183–196, 2016.
4. Pablo Juliano, Sandra Temmel, Manoj Rout, Piotr Swiergon, Raymond Mawson, and Kai Knoerzer. Creaming enhancement in a liter scale ultrasonic reactor at selected transducer configurations and frequencies. *Ultrasonics sonochemistry*, 20(1):52–62, 2013.
5. John J Milledge and Sonia Heaven. A review of the harvesting of micro-algae for biofuel production. *Reviews in Environmental Science and Bio/Technology*, 12:165–178, 2013.
6. Wolfgang Fraunhofer and Gerhard Winter. The use of asymmetrical flow field-flow fractionation in pharmaceutics and biopharmaceutics. *European journal of pharmaceutics and biopharmaceutics*, 58(2):369–383, 2004.
7. Swee Pin Yeap, JitKang Lim, Boon Seng Ooi, and Abdul Latif Ahmad. Agglomeration, colloidal stability, and magnetic separation of magnetic nanoparticles: collective influences on environmental engineering applications. *Journal of Nanoparticle Research*, 19:1–15, 2017.
8. Rohollah Nasiri, Amir Shamloo, Samad Ahadian, Leyla Amirifar, Javad Akbari, Marcus J Goudie, KangJu Lee, Nureddin Ashammakhi, Mehmet R Dokmeci, Dino Di Carlo, et al. Microfluidic-based approaches in targeted cell/particle separation based on physical properties: fundamentals and applications. *Small*, 16(29):2000171, 2020.
9. P Sajeesh and Ashis Kumar Sen. Particle separation and sorting in microfluidic devices: a review. *Microfluidics and nanofluidics*, 17:1–52, 2014.
10. J McGrath, M Jimenez, and H Bridle. Deterministic lateral displacement for particle separation: a review. *Lab on a Chip*, 14(21):4139–4158, 2014.
11. Dino Di Carlo. Inertial microfluidics. *Lab on a Chip*, 9(21):3038–3046, 2009.
12. Benjamin Abécassis, Cécile Cottin-Bizonne, C Ybert, A Ajdari, and L Bocquet. Boosting migration of large particles by solute contrasts. *Nature materials*, 7(10):785–789, 2008.
13. Benjamin Abécassis, Cécile Cottin-Bizonne, Christophe Ybert, Armand Ajdari, and Lydéric Bocquet. Osmotic manipulation of particles for microfluidic applications. *New Journal of Physics*, 11(7):075022, 2009.
14. Sangwoo Shin, Orest Shardt, Patrick B Warren, and Howard A Stone. Membraneless water filtration using $CO_2$. *Nature communications*, 8(1):15181, 2017.
15. Kyunghun Lee, Jongwan Lee, Dogyeong Ha, Minseok Kim, and Taesung Kim. Low-electric-potential-assisted diffusiophoresis for continuous separation of nanoparticles on a chip. *Lab on a Chip*, 20(15):2735–2747, 2020.
16. Trevor J Shimokusu, Vanessa G Maybruck, Jesse T Ault, and Sangwoo Shin. Colloid separation by $CO_2$-induced diffusiophoresis. *Langmuir*, 36(25):7032–7038, 2019.
17. Myungjin Seo, Sungmin Park, Dokeun Lee, Hyomin Lee, and Sung Jae Kim. Continuous and spontaneous nanoparticle separation by diffusiophoresis. *Lab on a Chip*, 20(22):4118–4127, 2020.
18. Adnan Chakra, Naval Singh, Goran T Vladisavljevic, François Nadal, Cécile Cottin-Bizonne, Christophe Pirat, and Guido Bolognesi. Continuous manipulation and characterization of colloidal beads and liposomes via diffusiophoresis in single-and double-junction microchannels. *ACS nano*, 17(15):14644–14657, 2023.
19. Brian J Kirby and Ernest F Hasselbrink Jr. Zeta potential of microfluidic substrates: 1. theory, experimental techniques, and effects on separations. *Electrophoresis*, 25(2):187–202, 2004.
20. Brian J Kirby and Ernest F Hasselbrink Jr. Zeta potential of microfluidic substrates: 2. data for polymers. *Electrophoresis*, 25(2):203–213, 2004.
21. Saebom Lee, Jinkee Lee, and Jesse T Ault. The role of variable zeta potential on diffusiophoretic and diffusioosmotic transport. *Colloids and Surfaces A: Physicochemical and Engineering Aspects*, 659:130775, 2023.
22. Jesse T Ault, Patrick B Warren, Sangwoo Shin, and Howard A Stone. Diffusiophoresis in one-dimensional solute gradients. *Soft matter*, 13(47):9015–9023, 2017.
23. Sangwoo Shin, Jesse T Ault, Jie Feng, Patrick B Warren, and Howard A Stone. Low-cost zeta potentiometry using solute gradients. *Advanced Materials*, 29(30):1701516, 2017.
24. Sangwoo Shin, Eujin Um, Benedikt Sabass, Jesse T Ault, Mohammad Rahimi, Patrick B Warren, and Howard A Stone. Size-dependent control of colloid transport via solute gradients in dead-end channels. *Proceedings of the National Academy of Sciences*, 113(2):257–261, 2016.
25. Jérémie Palacci, Benjamin Abécassis, Cécile Cottin-Bizonne, Christophe Ybert, and Lydéric Bocquet. Colloidal motility and pattern formation under rectified diffusiophoresis. *Physical review letters*, 104(13):138302, 2010.
26. Naval Singh, Goran T Vladisavljevic, François Nadal, Cécile Cottin-Bizonne, Christophe Pirat, and Guido Bolognesi. Enhanced accumulation of colloidal particles in microgrooved channels via diffusiophoresis and steady-state electrolyte flows. *Langmuir*, 38(46):14053–14062, 2022.
27. Naval Singh, Goran T Vladisavljević, François Nadal, Cécile Cottin-Bizonne, Christophe Pirat, and Guido Bolognesi. Reversible trapping of colloids in microgrooved channels via diffusiophoresis under steady-state solute gradients. *Phys. Rev. Lett.*, 125(24):248002, 2020.
28. Naval Singh, Adnan Chakra, Goran T Vladisavljević, Cécile Cottin-Bizonne, Christophe Pirat, and Guido Bolognesi. Composite norland optical adhesive (noa)/silicon flow focusing devices for colloidal particle manipulation and synthesis. *Colloids Surf., A*, 652:129808, 2022.
29. Martin K Rasmussen, Jonas N Pedersen, and Rodolphe Marie. Size and surface charge characterization of nanoparticles with a salt gradient. *Nature communications*, 11(1):2337, 2020.
30. Aura Visan and Rob GH Lammertink. Reaction induced diffusio-phoresis of ordinary catalytic particles. *Reaction Chemistry & Engineering*, 4(8):1439–1446, 2019.
31. Burak Akdeniz, Jeffery A Wood, and Rob GH Lammertink. Diffusiophoresis and diffusio-osmosis into a dead-end channel: Role of the concentration-dependence of zeta potential. *Langmuir*, 39(6):2322–2332, 2023.
32. Burak Akdeniz, Jeffery A Wood, and Rob GH Lammertink. Diffusiophoretic behavior of polyelectrolyte-coated particles. *Langmuir*, 40(11):5934–5944, 2024.
33. Lara R Lechlitner and Onofrio Annunziata. Macromolecule diffusiophoresis induced by concentration gradients of aqueous osmolytes. *Langmuir*, 34(32):9525–9531, 2018.
34. Suin Shim, Sepideh Khodaparast, Ching-Yao Lai, Jing Yan, Jesse T Ault, Bhargav Rallabandi, Orest Shardt, and Howard A Stone. Co 2-driven diffusiophoresis for maintaining a bacteria-free surface. *Soft Matter*, 17:2568–2576, 2021.
35. Robben E. Migacz, Guillaume Durey, and Jesse T. Ault. Convection rolls and three-dimensional particle dynamics in merging solute streams. *Phys. Rev. Fluids*, 8:114201, Nov 2023. doi: 10.1103/PhysRevFluids.8.114201.
36. CF Zukoski IV and DA Saville. An experimental test of electrokinetic theory using measurements of electrophoretic mobility and electrical conductivity. *Journal of colloid and interface science*, 107(2):322–333, 1985.
37. A Fernández Barbero, R Martínez García, MA Cabrerizo Vílchez, and R Hidalgo-Alvarez. Effect of surface charge density on the electrosurface properties of positively charged polystyrene beads. *Colloids and Surfaces A: Physicochemical and Engineering Aspects*, 92(1-2):121–126, 1994.
38. Jill E Seebergh and John C Berg. Evidence of a hairy layer at the surface of polystyrene latex particles. *Colloids and Surfaces A: Physicochemical and Engineering Aspects*, 100:139–153, 1995.
39. Marcel Minor, Ab J van der Linde, and Johannes Lyklema. Streaming potentials and conductivities of latex plugs in indifferent electrolytes. *Journal of colloid and interface science*, 203(1):177–188, 1998.
40. J Lyklema and M Minor. On surface conduction and its role in electrokinetics. *Colloids and Surfaces A: Physicochemical and Engineering Aspects*, 140(1-3):33–41, 1998.
41. Irina Ermolina and Hywel Morgan. The electrokinetic properties of latex particles: comparison of electrophoresis and dielectrophoresis. *Journal of colloid and interface science*, 285(1):419–428, 2005.
42. Songhua Lu, Kairuo Zhu, Wencheng Song, Gang Song, Diyun Chen, Tasawar Hayat, Njud S Alharbi, Changlun Chen, and Yubing Sun. Impact of water chemistry on surface charge and aggregation of polystyrene microspheres suspensions. *Science of the total environment*, 630:951–959, 2018.
43. Maryna V Manilo, Nikolai I Lebovka, and Sandor Barany. Effects of sort and concentration of salts on the electrosurface properties of aqueous suspensions containing hydrophobic and hydrophilic particles: Validity of the hofmeister series. *Journal of Molecular Liquids*, 276:875–884, 2019.
44. Philippe Leroy, Nicolas Devau, André Revil, and Mohamed Bizi. Influence of surface conductivity on the apparent zeta potential of amorphous silica nanoparticles. *Journal of colloid and interface science*, 410:81–93, 2013.
45. Robert J Hunter. The significance of stagnant layer conduction in electrokinetics. *Advances in colloid and interface science*, 100:153–167, 2003.
46. Gwo-Bin Lee, Chih-Chang Chang, Sung-Bin Huang, and Ruey-Jen Yang. The hydrodynamic focusing effect inside rectangular microchannels. *Journal of Micromechanics and Microengineering*, 16(5):1024, 2006.
47. Ankur Gupta, Suin Shim, and Howard A Stone. Diffusiophoresis: from dilute to concentrated electrolytes. *Soft Matter*, 16(30):6975–6984, 2020.
48. Dennis C Prieve, Stephanie M Malone, Aditya S Khair, Robert F Stout, and Mazen Y Kanj. Diffusiophoresis of charged colloidal particles in the limit of very high salinity. *Proceedings of the National Academy of Sciences*, 116(37):18257–18262, 2019.
49. Marek Bekir, Marcel Sperling, Daniela Vasquez Muñoz, Cevin Braksch, Alexander Böker, Nino Lomadze, Mihail N Popescu, and Svetlana Santer. Versatile microfluidics separation of colloids by combining external flow with light-induced chemical activity. *Advanced Materials*, 35(25):2300358, 2023.
50. Jerome FL Duval, Jose Paulo S Farinha, and Jose P Pinheiro. Impact of electrostatics on the chemodynamics of highly charged metal–polymer nanoparticle complexes. *Langmuir*, 29(45):13821–13835, 2013.
51. Partha P Gopmandal and Jérôme FL Duval. Electrostatics and electrophoresis of engineered nanoparticles and particulate environmental contaminants: beyond zeta potential-based formulation. *Current Opinion in Colloid & Interface Science*, 60:101605, 2022.
52. John L Anderson. Colloid transport by interfacial forces. *Annual review of fluid mechanics*, 21(1):61–99, 1989.
53. Hiroyuki Ohshima. A simple expression for henry's function for the retardation effect in electrophoresis of spherical colloidal particles. *Journal of Colloid and Interface Science*, 168(1):269–271, 1994.
54. Huan J Keh and Yeu K Wei. Diffusiophoretic mobility of spherical particles at low potential and arbitrary double-layer thickness. *Langmuir*, 16(12):5289–5294, 2000.
55. Jacob Heskel Masliyah. *Electrokinetic transport phenomena*. Edmonton, Alta.: Alberta Oil Sands Technology and Research Authority, Alberta, Canada, 1994.
56. Min Gu. *Advanced Optical Imaging Theory*, volume 75 of *Springer Series in Optical Sci-*

*ences*. Springer Berlin: Heidelberg, 1 edition, 1999.
57. David A Agard. Optical sectioning microscopy: cellular architecture in three dimensions. *Annual review of biophysics and bioengineering*, 13(1):191–219, 1984.
58. Naval Singh, Goran T. Vladisavljević, Fran çois Nadal, Cécile Cottin-Bizonne, Christophe Pirat, and Guido Bolognesi. Reversible trapping of colloids in microgrooved channels via diffusiophoresis under steady-state solute gradients. *Phys. Rev. Lett.*, 125:248002, Dec 2020. doi: 10.1103/PhysRevLett.125.248002.
59. ISO 22412:2017. Particle size analysis — Dynamic light scattering (DLS). Standard, International Organization for Standardization, Geneva, CH, February 2017.

# Supplementary Materials for
# Surface Chemistry-based Continuous Separation of Colloidal Particles via Diffusiophoresis and Diffusioosmosis

Adnan Chakra, Christina Puijk, Goran T. Vladisavljević,
Cécile Cottin-Bizonne, Christophe Pirat, Guido Bolognesi*
*Corresponding author. Email: g.bolognesi@ucl.ac.uk

**This PDF file includes:.** Supplementary Text
Figs. S1 to S9
Tables S1 to S3

## Particle Focusing Mechanism

The formation of the peaks in the double-junction microfluidic device, described in Figure 1 of the main text, is driven by the vertical $y$-axis component of the salt concentration gradient, generated by the Poiseuille-like velocity profile in the rectangular channel. Indeed, as shown by the numerical simulation results (Figure 3 of the main text), the parabolic velocity profile induces a slight curvature to the salt concentration isolines, which translates into a vertical component of the salt concentration gradient and, thus, of the diffusiophoresis velocity. In the inner region of the channel ($|x|/w_m < 0.5$), this component is directed from the bulk to the channel walls, leading to the accumulation of particles at the top and bottom walls. Once particle peaks form in close proximity to the wall, their transversal motion is governed by two competing factors, as shown in Figure S1. One is particle diffusiophoresis in the $x$ direction, $u_{DP,x}$, pointing outwards. The other one is a counter-acting convective flow, $u_x$, pointing inwards, which is induced by the diffusioosmotic slip velocity at the horizontal walls of the channel, $u_{DO,x} = -\Gamma_{DO}\frac{\partial \ln c}{\partial x}$, with $\Gamma_{DO}$ the diffusiophoresis mobility. The total transversal particle velocity is given by $u_{p,x} = u_x + u_{DP,x}$ and, close to the wall (where $u_x \simeq u_{DO,x}$), it can be re-expressed as:

$$u_{p,x} = \Gamma_{\text{eff}}\frac{\partial \ln c}{\partial x} \tag{S1}$$

where $\Gamma_{\text{eff}} = \Gamma_{DP} - \Gamma_{DO}$ is denoted as the effective mobility (35). Hence, when $\Gamma_{\text{eff}} > 0$ (namely, diffusiophoresis is dominant and $\Gamma_{DP} > \Gamma_{DO}$), $u_{p,x}$ has the same direction of the salt concentration gradient, and the particles accumulated near the wall, migrate towards the outer regions of the channel. Conversely, when $\Gamma_{\text{eff}} < 0$ (namely, diffusioosmosis is dominant and $\Gamma_{DP} < \Gamma_{DO}$), $u_{p,x}$ has opposite direction to the salt concentration gradient, and particles migrate towards the central region of the channel. Figure S4 shows the calculated values of particle effective mobility $\Gamma_{\text{eff}}$ over a range of LiCl concentrations for all types of particles investigated in this study.

## Microfluidic devices

In this study, we used double Ψ- junction microfluidic chips, with two different variations. All devices had a nominal depth of $h = 45$ μm. The first variation of the double Ψ- junction microchip was designed with a single outlet and was used in the investigations on the effect of ionic strength on particle dynamics. The inner inlet of the upstream junction had a width of $w_i = 100\,\mu$m, while the downstream junction had a middle channel width of $w_m = 250\,\mu$m. The width of the side channels in the upstream and downstream junctions were $75\,\mu$m each. This device had a total length of $z/w_m = 80$ (20 mm). The second version of the $\Psi$-junction microchip was used for the separation experiments. This device had three outlets and revised dimensions for the upstream and downstream junctions. For the upstream junction, the width of the inner inlet channel was $w_i = 100\,\mu$m. For the downstream junction, the width of the middle channel was $w_m = 200\,\mu$m. The width of the side channels in both upstream and downstream junctions were $50\,\mu$m each. Finally, the width of the middle outlet was $75\,\mu$m, while the width of the side outlets were $112.5\,\mu$m each. This device had a total length of $z/w_m = 200$ (40 mm).

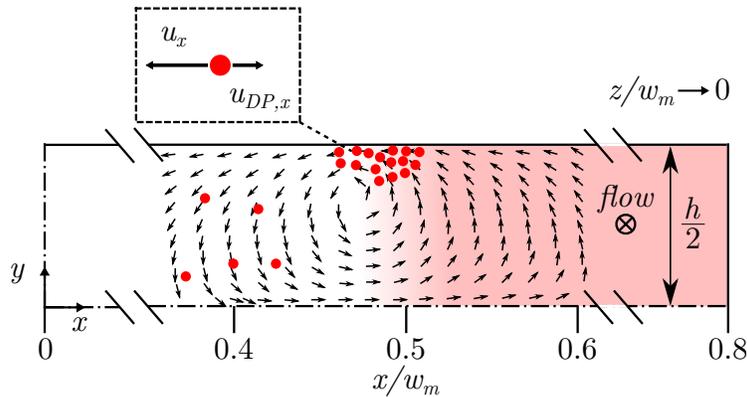

**Fig. S1. Focusing Mechanism.** Schematic of the top right quarter of the $x$-$y$ cross-section of the device, perpendicular to the flow direction, illustrating particle dynamics close to the downstream junction ($z/w_m \to 0$) under the influence of a salt gradient. Particle peaks, forming nearby where the colloidal and higher salinity stream meet ($x/w_m \simeq 0.5$), converge inwards when $u_x > u_{DP_x}$, namely $\Gamma_{\text{eff}} < 0$. Arrows exemplify the hydrodynamic convection rolls induced by the slip velocity $u_{DO_x}$ at the wall.

## Diffusioosmosis mobility of glass and PDMS walls

The diffusioosmosis (DO) mobility coefficient $\Gamma_{\text{DO}}$ is calculated as (52)

$$\Gamma_{\text{DO}} = \frac{\varepsilon}{2\eta}\left(\frac{k_bT}{Ze}\right)^2 \left[2\beta\bar{\zeta} - 4\ln(1-\gamma^2)\right] \quad \text{(S2)}$$

where $\gamma = \tanh\left(\bar{\zeta}/4\right)$ and $\beta = \frac{D_+ - D_-}{D_+ + D_-}$, $\varepsilon$ is the absolute permittivity of the surrounding liquid medium, $\eta$ is the medium viscosity, $k_bT$ is the thermal energy, $Z$ is the ion valence, $e$ is the elementary charge, $\bar{\zeta} = \frac{Ze\zeta}{k_bT}$ is the adimensionalized zeta potential and $D_+$ and $D_-$ are the diffusivities of cations and anions, respectively. For LiCl aqueous solution at 25°C, $\varepsilon = 85.8$, $\eta = 0.9\,\text{Pa}\,\text{s}$, $D_+ = 1.026 \times 10^{-9}\,\text{m}^2/\text{s}$ and $D_- = 1.964 \times 10^{-9}\,\text{m}^2/\text{s}$. The empirical model developed by Kirby and Hasselbrink Jr (19, 20) was employed to estimate the zeta potential of the glass and PDMS channel walls. The model applies to indifferent univalent electrolytes for which the surface-ion binding is independent on the ionic strength of the solution. By assuming that the surface charge density does not depend on the counterion type and concentration, the model predicts a logarithmic scaling of the zeta potential with the salt concentration in the high zeta potential limit ($\zeta \gg 2k_bT/e$). The model was validated against comprehensive data sets obtained from literature, and it was showed how it can predict accurately the zeta potential for silica and different polymer substrates across various salt concentrations and pH levels. Although the model was validated against zeta potential data sets with Na$^+$ and K$^+$ counterions, it is expected that the model assumptions are still valid for Li counterions. Therefore, the zeta potential of the glass and PDMS walls in our microfluidic devices can be approximated as

$$\zeta = a_1 \log_{10} c^* \quad \text{(S3)}$$

where $c = \sum_i c_i$ is the sum of the concentration of all counterion species in the solution, $c^* = c/1\text{M}$ is the dimensionless total counterion concentration, and $a_1$ is the correlation coefficient. Note that for the range of salt concentrations examined in this study (0.01-100 mM), the concentration of H$^+$ counterions at neutral conditions is negligible compared to the concentration on Li$^+$ counterions. Therefore the term $c$ in Eq. S3 can be assumed equal to the concentration of LiCl salt in the solution. For silica substrates, it was found that $a_1 \simeq 2 + 7 \cdot (\text{pH} - 3)$, thus, $a_1 = -30\,\text{mV}$ at neutral conditions (pH=7). For PDMS substrates exposed to an electrolyte solution in the pH range between 6.5 and 7, the fit of the model to the experimental data provides a value of the correlation coefficient $a_1$ of ca. $-27\,\text{mV}$. The accuracy of this model was also recently confirmed by streaming potential measurements for PDMS surfaces exposed to NaCl salt concentrations in the range from 0.1 to 10 mM (31). In the numerical simulations, the zeta potential of the channel walls was calculated through Eq. S3 with $a_1 = -27\,\text{mV}$. The resulting diffusioosmosis mobility, calculated according to Eq. S2, is plotted as a function of salt concentration in Figure S3.

## 1. Diffusiophoresis mobility of colloidal particles

Particle size and zeta potential were measured via dynamic light scattering and electrophoretic light scattering for LiCl concentrations within the range from 0.1 to 100 mM. Note that zeta potential measurements at a salt concentration of 0.01 mM were not reproducible, possibly due the uncontrolled presence of ionic contaminants in the sample solutions at micro-molar concentrations. The instrument provides zeta potential values based on Smoluchowski's module, assuming the Debye length is negligible compared to the particle size. To account for the finite size of the electric double layer, the zeta potentials were corrected using the Henry's model with Ohshima's approximated analytical expression for Henry's function (53). The measured diameters for all particles species examined in this study are reported in Table S1. The particle diameters were calculated as the average over multiple dynamic light scattering measurements at varying salt concentrations ($n \geq 6$) within the range from 0.1 to 100 mM, whereas the uncertainties on the diameters were calculated as the standard error of the mean. Figure S2 shows the zeta potential measurements under varying salt concentrations for the different particles. The zeta potentials of 0.02, 0.1, 0.2, 0.5 µm red particles exhibit a monotonic trend and, in accordance with a previous study (31), the following relation

$$\zeta = a_0 + a_1 \log_{10} c^* \tag{S4}$$

was used to fit the experimental data. The coefficients $a_0$ and $a_1$ are determined by fitting Eq. S4 to the zeta potential measurements and their values are reported in Table S2.

Table S1. Particle diameter measured by dynamic light scattering and particle carboxyl content.

| Nominal Diameter (nm) | Color | Diameter $d_{DLS}$ (nm) | Specific carboxyl equivalent (meq/g) | Specific surface area ($10^5\,\text{cm}^2/\text{g}$) | Surface carboxyl concentration ($10^{-10}\,\text{mol/cm}^2$) |
|---|---|---|---|---|---|
| 20 | red | 38 ± 2 | 0.515 | 22 | 2.34 |
| 100 | red | 118 ± 1 | 0.31 | 5.2 | 5.96 |
| 200 | red | 226 ± 1 | 0.3772 | 2.7 | 14.0 |
| 500 | red | 539 ± 5 | 0.0124 | 1.2 | 1.03 |
| 500 | yellow-green | 509 ± 3 | 0.1077 | 1.2 | 8.98 |
| 1000 | red | 1181 ± 34 | 0.0247 | 0.57 | 4.33 |

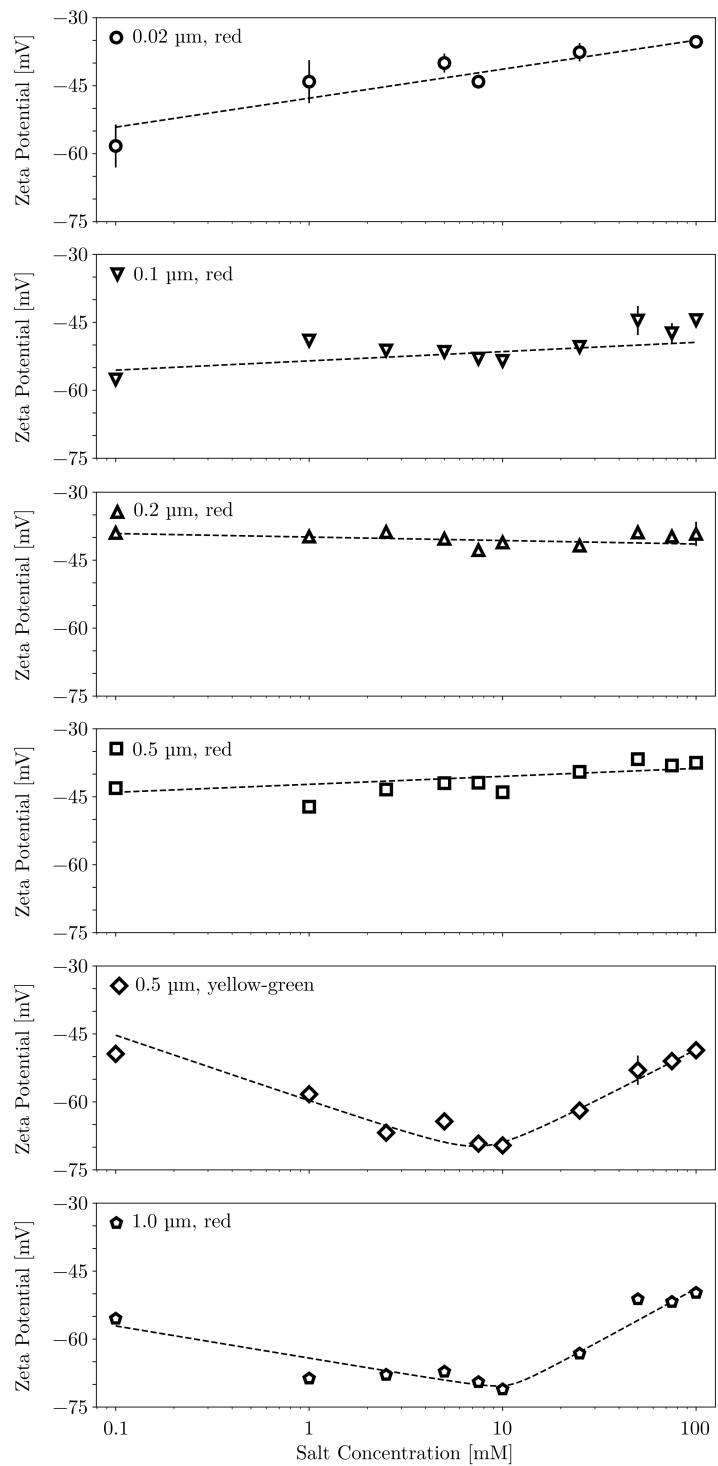

**Fig. S2. Zeta potential as a function of LiCl salt concentration for various types of colloidal particles.** Markers are experimental data and dashed solid lines are the fits to Eq. S4 for fluorescent red sub-micron particles and to Eq. S5 for the $1\,\mu m$ fluorescent red and $0.5\,\mu m$ fluorescent yellow-green particles. The values of the best-fitting coefficients are reported in Tables S2 and S3. Error bars are calculated as the standard deviation of three repeated measurements of a sample.

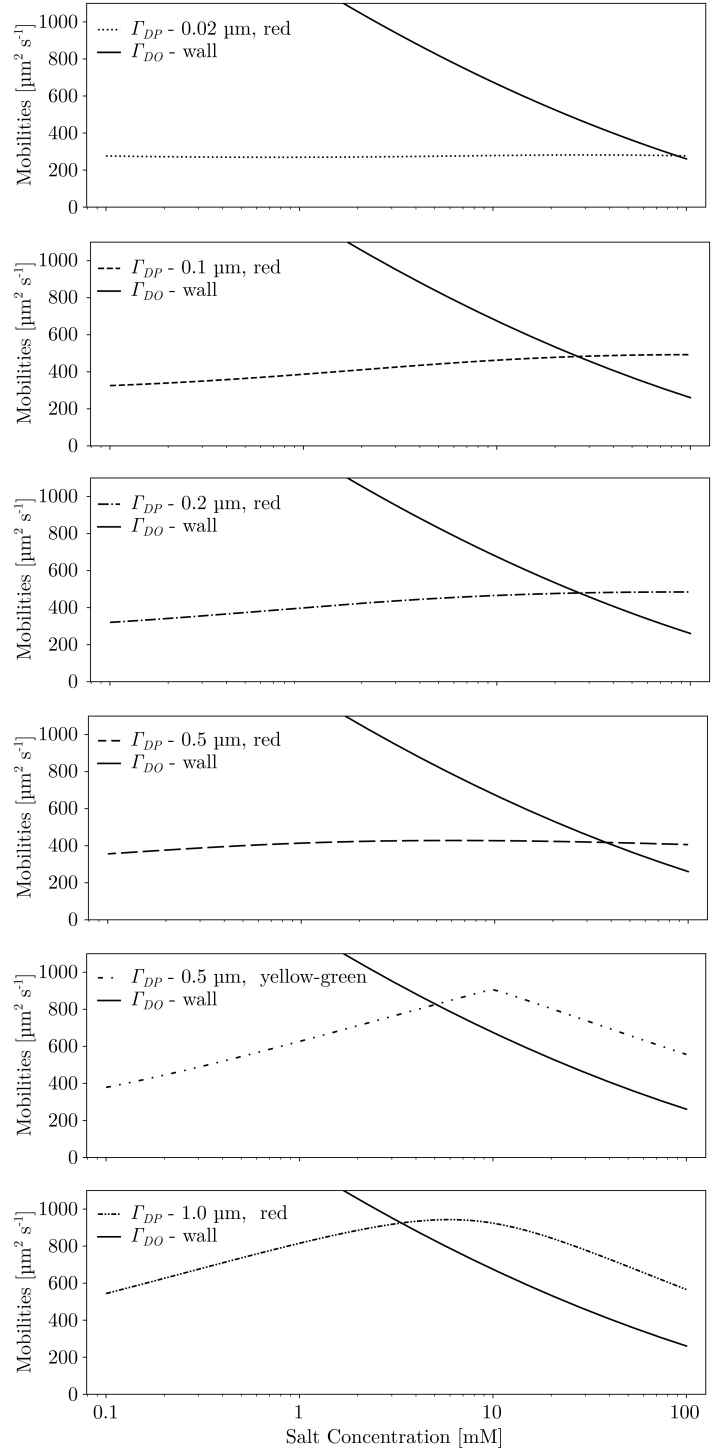

**Fig. S3. Diffusiophoresis and diffusioosmosis mobilities as a function of LiCl salt concentration for various types of colloidal particles.** Broken lines are the diffusiophoresis mobilities of various types of colloidal particles calculated via Eq. S4, (S5), and (S6) with the coefficients $a_i, (i = 0, \ldots, 4)$ reported in Tables S2 and S3. Solid lines are the diffusioosmosis mobility of the channel walls calculated via Eq. S2 with $a_1 = -27\,\text{mV}$.

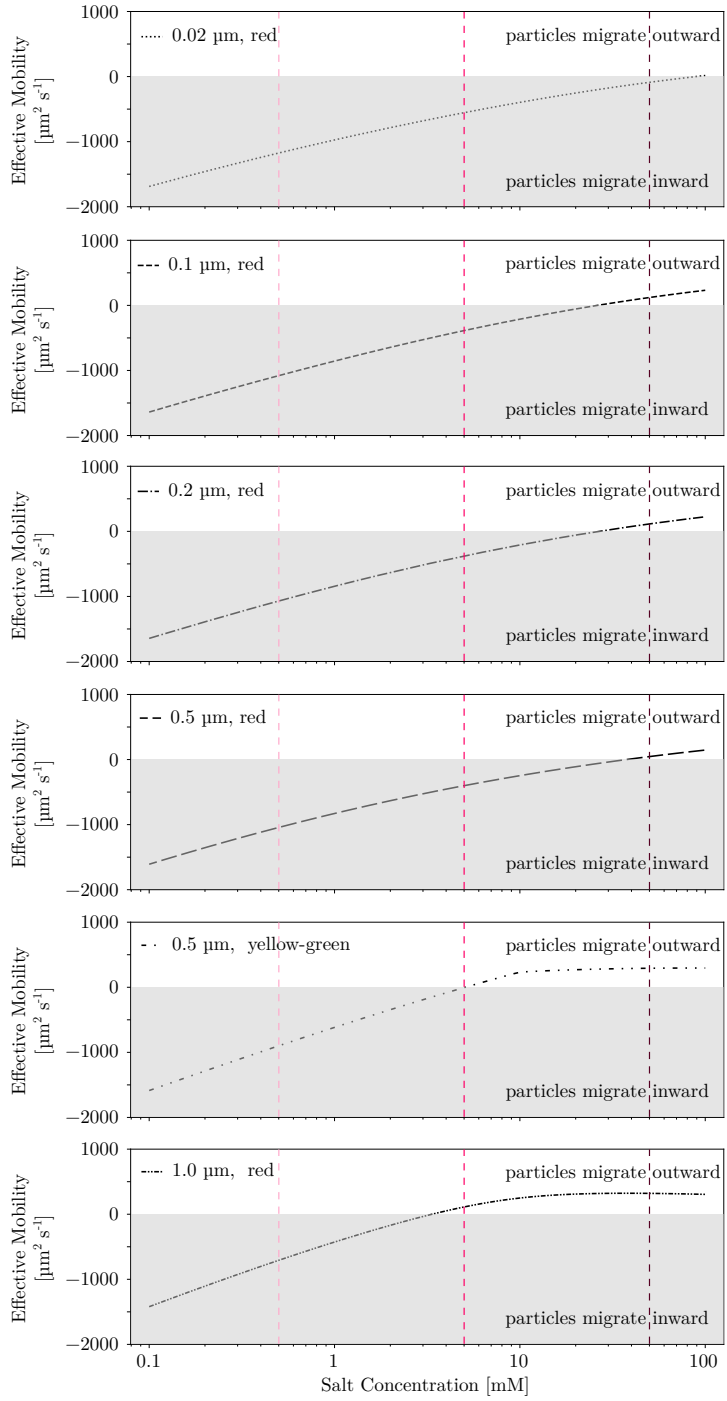

**Fig. S4. Effective mobilities $\Gamma_{\text{eff}} = \Gamma_{\text{DP}} - \Gamma_{\text{DO}}$ as a function of LiCl salt concentration for various types of colloidal particles.** The vertical dashed lines corresponds to the average salt concentration $(c_L + c_H)/2$ for low (light red), moderate (red) and high (dark red) ionic strength conditions, namely $0.5\,\text{mM}$, $5\,\text{mM}$ and $50\,\text{mM}$.

Conversely, 0.5 µm yellow-green and 1.0 µm red particles display a non-monotonic dependence of the zeta potential on salt concentration. Migacz et al. (35) reported a similar trend for 0.2 µm carboxylate polystyrene particles and they adopted the following relation

$$\zeta = a_0 + a_1 \ln c^* + \sqrt{a_2 + a_3 \ln c^* + a_4 (\ln c^*)^2} \tag{S5}$$

to fit their dataset. This relation was used to fit the zeta potential measurements of the 0.5 µm yellow-green and 1.0 µm red particles, and the best-fitting coefficients $a_i, (i = 0, \ldots, 4)$ are reported in Table S3.

The specific carboxyl equivalent $c_{eq}$, defined as the chemical equivalent (eq) of the reactive carboxyl groups per unit mass of particles, and the specific surface area $S$ of all particles are reported in Table S1. This data were sourced from the certificates of analysis of the lot numbers of the examined particles, available on the supplier manufacturer website. The surface concentration of the reactive surface carboxyl group was estimated as $c_{eq}/S$ and it is reported in Table S1.

Using the semi-analytical model developed by Keh and Wei et al. (54), the diffusiophoresis mobility is calculated as

$$\Gamma_{\text{DP}} = \frac{\varepsilon}{\eta} \left[ \frac{k_b T}{Ze} \Theta_1(\lambda) \beta \zeta + \frac{1}{8} \Theta_2(\lambda) \zeta^2 + \mathcal{O}(\zeta^3) \right] \tag{S6}$$

This expression is valid for any value of the ratio $\lambda = (\kappa a)^{-1}$ between the electric double layer thickness $\kappa^{-1}$ and the particle radius $a$. The following relations for the functions $\Theta_1(\lambda)$ and $\Theta_2(\lambda)$ are used (55)

$$\begin{aligned}\Theta_1(\lambda) &= 1 - \frac{1}{3}\left(1 + 0.07234 \lambda^{-1.129}\right)^{-1} \\ \Theta_2(\lambda) &= 1 - \left(1 + 0.085 \lambda^{-1} + 0.02 \lambda^{-0.1}\right)^{-1}\end{aligned} \tag{S7}$$

The calculated diffusiophoresis mobilities are plotted in Figure S3 as a function of salt concentration for all particle types examined in this study. The effective mobilities, calculated as $\Gamma_{\text{eff}} = \Gamma_{\text{DP}} - \Gamma_{\text{DO}}$, are shown in Figure S4.

## Normalization of fluorescence intensity profiles

For each microfluidic experiment on particle dynamics under a salinity gradient conducted in this study, a corresponding control experiment (without salt concentration gradient) was also performed. Image acquisition settings, including location of the focal plane, exposure time and gain parameter, were identical in both the experiment and the corresponding control. It is worth noting that the epi-fluorescence micrographs are generated from the convolution of the particle fluorescence intensity with the microscope point-spread function (56). Thus, the micrographs are the result of an integration of the particle fluorescence intensity over an optical window, whose characteristic size in the vertical $y$-direction is of the order of the depth of field of the optical system (ca. 10 µm). Note that the actual depth of this optical window is larger than the depth of field, since the fluorescence signal generated by out-of-focus particles also contribute to the formation of the image at the camera sensor. The raw epi-fluorescence micrographs were cropped in the central region of the image. The fluorescence intensity $I_{image}(x,z)$ of the resulting rectangular cropped image was averaged along the flow direction $z$ to obtain the fluorescence intensity profile $I(x) = 1/w_z \int I_{image}(x,z) \mathrm{d}z$ with $w_z$ the size of the cropped image along the $z$ direction. The fluorescence intensity profiles were normalized with respect to the absolute fluorescence intensity values measured in the control experiments, according to the following procedure. Initially, a background noise intensity level, $I_{bg}$, was calculated as the minimum intensity value of the fluorescence profiles obtained under no salinity gradient conditions, namely $I_{bg} = \min(I_c(x))$, where $I_c(x)$ is the transverse fluorescence intensity profile of the control experiment under no salt concentration gradients. Afterwards, a reference intensity $I_0$ is calculated from the profile $I_c(x)$ as follows

$$I_0 = \frac{1}{w_c} \int \left(I_c(x) - I_{bg}\right) dx \tag{S8}$$

where the integral extends over the width $w_c$ of the colloidal stream. In other words, the reference intensity $I_0$ is the average value of the fluorescence intensity profile of the colloidal stream in the control experiment. Note that in control experiments, particles are homogeneously distributed across the channel depth, as showed by confocal microscopy analysis in our previous study (18). Subsequently, the normalized transverse fluorescence intensity profile $I_{norm}(x)$ was calculated as:

$$I_{norm}(x) = \frac{I(x) - I_{bg}}{I_0} \tag{S9}$$

**Table S2. Values of the best-fitting coefficients calculated by fitting Eq. S4 to the zeta potential measurements shown in Figure S2.**

| Particle Diameter (µm) | Fluorescence Color | $a_0$/[mV] | $a_1$/[mV] |
|---|---|---|---|
| 0.02 | red | $-28.5 \pm 2.0$ | $6.40 \pm 1.0$ |
| 0.1 | red | $-47.4 \pm 0.9$ | $2.0 \pm 0.4$ |
| 0.2 | red | $-44.1 \pm 0.6$ | $1.5 \pm 0.2$ |
| 0.5 | red | $-37.0 \pm 0.4$ | $1.8 \pm 0.2$ |

**Table S3. Values of the best-fitting coefficients calculated by fitting Eq. S5 to the zeta potential measurements shown in Figure S2.**

| Particle Diameter (μm) | Fluorescence Color | $a_0$ [mV] | $a_1$ [mV] | $a_2$ [mV$^2$] | $a_3$ [mV$^2$] | $a_4$ [mV$^2$] |
|---|---|---|---|---|---|---|
| 0.5 | yellow-green | $-65.2\pm1.6$ | $1.6\pm0.6$ | $1500\pm300$ | $620\pm110$ | $64\pm13$ |
| 1.0 | red | $-55.2\pm1.5$ | $3.5\pm0.3$ | $910\pm160$ | $400\pm160$ | $44\pm8$ |

Figure S5 presents a calibration curve for the average fluorescence intensity measured for samples with varying particle concentrations, $n$. Four separate experiments were conducted, each testing a sample with a different colloidal concentration ($n = 0.025$ %, 0.05 %, 0.075 %, and 0.125 % v/v) injected inside a $\Psi$-junction device. In each experiment, the colloidal solution was injected into all the inlet channels of the device. Micrographs were acquired at a given distance downstream of the junction. Each image was cropped into a rectangular region of size $w_x \times w_z$ and the average fluorescence intensity of the cropped region was calculated as $I_{av} = (w_x w_z)^{-1} \iint I_{image}(x,z) \mathrm{d}x \mathrm{d}z$.

The dataset $I_{av}$ vs $n$ is nicely fitted by a straight line, establishing a direct linear relationship between measured fluorescence intensity and colloidal concentration of the solutions flowing inside the device. This observation ensures that the fluorescence intensity emitted by the particles is proportional to the particle concentration. Consequently, the measured fluorescence intensity profile can be expressed as the convolution of the particle concentration field $n(x,y,z)$ with the microscope point-spread function $s(x,y,z)$

$$I(x) = \beta \bar{n}(x) + I_{\text{bg}} \tag{S10}$$

with

$$\bar{n}(x) = \int \frac{\mathrm{d}z}{w_z} \iiint n(x',y',z') s(x-x', y_0-y', z-z') \mathrm{d}x' \mathrm{d}y' \mathrm{d}z' \tag{S11}$$

where $y_0$ is the location of the focal plane and $\beta$ a constant of proportionality (57). The microscope point-spread function is normalized so that $\iiint s(x,y,z) \mathrm{d}x\,\mathrm{d}y\,\mathrm{d}z = 1$. Assuming a constant particle concentration $n$ and averaging over the $x$ direction, Eq. S10 gives $I_{av} = \beta n + I_{\text{bg}}$, which is in agreement with the linear trend observed for the calibration curve. Similarly, in control experiments, where $n(x,y,z) = n_0$ for locations $x$ within the colloidal streams, Eq. S10 gives $I_c(x) = \beta n_0 + I_{\text{bg}}$, which combined with Eq. S8 leads to $I_0 = \beta n_0$. This relation together with Eq. S9 and Eq. S10 leads to

$$I_{norm}(x) = \frac{\bar{n}(x)}{n_0} \tag{S12}$$

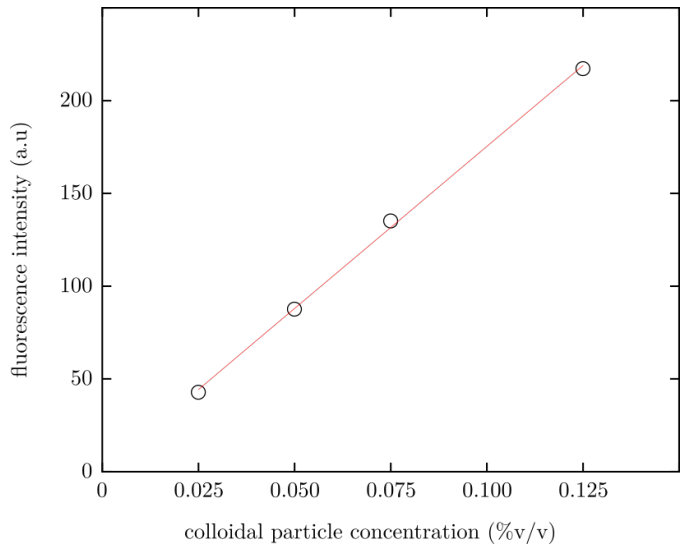

**Fig. S5. Fluorescence intensity calibration curve.** Relationship between the average fluorescent intensity and the colloidal concentration of 200 nm red fluorescent carboxylate polystyrene particles.

Therefore, the normalized fluorescence intensity profile provides a measure of depth-averaged particle concentration profile normalized with respect to the concentration of the solution injected into the device. More specifically, $I_{\text{norm}}(x) = 0$ corresponds to locations $x$ with no particles ($\bar{n}(x) = 0$), whereas $I_{\text{norm}}(x) = 1$ corresponds to locations $x$ with a depth-average particle concentration close, but not necessarily identical, to the concentration of the colloidal solution injected into the device ($\bar{n}(x) = n_0$). Indeed, $\bar{n}(x)$ does not correspond exactly to the depth-average particle concentration, unless particles are homogeneously distributed across the depth of the channel as in control experiments. In presence of salinity gradients, the particle concentration varies significantly between the channel bulk and nearby the channel walls, thus the term $\bar{n}(x)$ differs from the depth-average particle concentration.

## Numerical model

The 3D computational domain consists in a rectangular channel of width $w$, depth $h$ and length $l$. The following set of dimensionless equations for the hydrodynamic velocity $\mathbf{u}$, pressure $p$, salt concentration $c$ and particle concentration $n$ are solved in COMSOL Multiphysics

$$\text{Re}\,\mathbf{u} \cdot \nabla \mathbf{u} = -\nabla p + \nabla^2 \mathbf{u} \tag{S13}$$

$$\nabla \cdot \mathbf{u} = 0 \tag{S14}$$

$$\text{Pe}_c\,\mathbf{u} \cdot \nabla c = \nabla^2 c \tag{S15}$$

$$\text{Pe}_n\,\nabla \cdot [(\mathbf{u} + \mathbf{u}_{\text{DP}})n] = \nabla^2 n \tag{S16}$$

All quantities are rescaled according to the following relations

$$u \propto U_0 \qquad x,y,z \propto w \qquad p \propto \frac{\eta U_0}{w} \qquad c \propto c_H \qquad n \propto n_0 \tag{S17}$$

where $U_0$ is the average hydrodynamic velocity along the flow direction, $w$ is the actual (dimensional) channel width, $\eta$ is the viscosity of both outer and inner solutions, $c_H$ is the solute concentration of the high salt concentration stream, $c_L$ is the solute concentration of the high salt concentration stream, $n_0$ is the particle concentration of the colloid solution injected into the device. The diffusiophoresis velocity is expressed as

$$\mathbf{u}_{\text{DP}} = \xi_{\text{DP}} \hat{\Gamma}_{\text{DP}}(c) \frac{\nabla c}{c} \tag{S18}$$

where $\hat{\Gamma}_{\text{DP}}(c)$ is a dimensionless diffusiophoresis mobility defined as $\hat{\Gamma}_{\text{DP}} = \Gamma_{\text{DP}}/\Gamma_{\text{DP,ref}}$. The diffusiophoresis mobility $\Gamma_{\text{DP}}$ is calculated as a function of the dimensional salt concentration, $c\,c_H$, as detailed in Section 1, whereas the reference value $\Gamma_{\text{DP,ref}}$ is arbitrarily chosen as the mobility at the low salt concentration, $\Gamma_{\text{DP,ref}} = \Gamma_{\text{DP}}(c_L)$. At the channel inlet, the boundary condition for the velocity field is $\mathbf{u} = \mathbf{u}_{\text{inlet}}$, with $\mathbf{u}_{\text{inlet}}$ the fully developed velocity field at a cross section of the rectangular channel perpendicular to the flow direction and with average velocity equal 1. The boundary conditions at channel inlet for the salt and particle concentration fields are

$$c(x,y,0) = \begin{cases} c_L/c_H & \text{if } |x| < \frac{w_m}{2w} \\ 1 & \text{if } |x| \geq \frac{w_m}{2w} \end{cases} \tag{S19}$$

$$n(x,y,0) = \begin{cases} 1 & \text{if } \frac{w_m - 2w_o}{2w} \leq |x| \leq \frac{w_m}{2w} \\ 0 & \text{if otherwise} \end{cases} \tag{S20}$$

where $w_o$ is the actual (dimensional) width of the side channels of the two junctions and $w_m$ is the actual (dimensional) width of the inner channel of the downstream junction. At the channel outlet, the zero normal gradient boundary condition for the pressure, salt and particle concentrations are imposed. At the remaining walls, the slip boundary condition

$$\mathbf{u} = -\xi_{\text{DO}} \hat{\Gamma}_{\text{DO}} \frac{\nabla c}{c} \tag{S21}$$

is applied together with the zero flux condition for the salt and particle concentration fields. $\hat{\Gamma}_{\text{DO}}(c)$ is a dimensionless diffusioosmosis mobility defined as $\hat{\Gamma}_{\text{DO}} = \Gamma_{\text{DO}}/\Gamma_{\text{DO,ref}}$. The diffusiophoresis mobility $\Gamma_{\text{DO}}$ is calculated as a function of the dimensional salt concentration, $c\,c_H$, as detailed in Section , whereas the reference value $\Gamma_{\text{DO,ref}}$ is arbitrarily chosen as the mobility at the low salt concentration, $\Gamma_{\text{DO,ref}} = \Gamma_{\text{DO}}(c_L)$. The channel outlet was located at 5 mm from the channel inlet plus 5 times the channel depth to ensure that the boundary conditions at the channel outlet do not affect the fields at the cross-section $z = 5$ mm. The dimensionless numbers, governing the examined system, are defined as follows

$$\text{Re} = \frac{\rho U_0 w}{\eta} \qquad \text{Pe}_c = \frac{w U_0}{D_s} \qquad \text{Pe}_n = \frac{w U_0}{D_p} \qquad \xi_{\text{DP}} = \frac{\Gamma_{\text{DP,ref}}}{w U_0} \qquad \xi_{\text{DO}} = \frac{\Gamma_{\text{DO,ref}}}{w U_0} \tag{S22}$$

where $\rho$ is the density of the inner and outer solutions and $D_p$ is the particle diffusivity, calculated as $k_b T/(6\pi\eta a)$. Despite in the experiments the channel walls are made of different materials (namely, glass for the bottom wall and PDMS for the remaining walls), in the simulation all walls have the same diffusioosmosis coefficient to take advantage of the symmetries of the problem, hence limiting the computational cost. As discussed in Section , glass and PDMS are expected to have similar diffusioosmosis mobilities. Additionally, the good agreement between the numerical simulations and experimental results suggests that this assumption is acceptable.

Instead of solving the coupled equations for the hydrodynamic velocity and salt concentration fields, we reformulated the problem by using a perturbation approach, as we did in a previous study (58). This method allows one to decouple the equations for the velocity and salt concentration fields, reducing the computer memory usage. The velocity, pressure and salt concentration fields are expressed as a power expansion of the parameter $\xi_{\text{DO}}$ as follows

$$\mathbf{u} = \mathbf{u}_0 + \xi_{\text{DO}}\,\mathbf{u}_1 + \mathcal{O}(\xi_{DO}^2) \tag{S23}$$

$$p = p_0 + \xi_{\text{DO}}\,p_1 + \mathcal{O}(\xi_{DO}^2) \tag{S24}$$

$$c = c_0 + \xi_{\text{DO}}\,c_1 + \mathcal{O}(\xi_{DO}^2) \tag{S25}$$

The $0^{th}$ and $1^{st}$ order terms of the fields are obtained by solving the following equations

$$\text{Re}\,\mathbf{u}_0 \cdot \nabla \mathbf{u}_0 = -\nabla p_0 + \nabla^2 \mathbf{u}_0 \tag{S26}$$

$$\nabla \cdot \mathbf{u_0} = 0 \tag{S27}$$

$$\text{Pe}_c\,\mathbf{u_0} \cdot \nabla c_0 = \nabla^2 c_0 \tag{S28}$$

$$\text{Re}\,(\mathbf{u}_1 \cdot \nabla \mathbf{u}_0 + \mathbf{u}_0 \cdot \nabla \mathbf{u}_1) = -\nabla p_1 + \nabla^2 \mathbf{u}_1 \tag{S29}$$

$$\nabla \cdot \mathbf{u_1} = 0 \tag{S30}$$

$$\text{Pe}_c\,(\mathbf{u_0} \cdot \nabla c_1 + \mathbf{u_1} \cdot \nabla c_0) = \nabla^2 c_1 \tag{S31}$$

The wall boundary conditions for the velocity fields are $u_0 = 0$ and

$$\mathbf{u}_1 = -\xi_{\text{DO}}\,\hat{\Gamma}_{\text{DP}}(c_0) \frac{\nabla c_0}{c_0} \tag{S32}$$

As a results, velocity and concentration fields can be now solved separately in the following order: $u_0$, $c_0$, $u_1$ and $c_1$. The particle concentration $n$ is determined at last by solving Eq.(S16) with the particle diffusiophoresis velocity expressed as

$$\mathbf{u}_{\text{DP}} = \xi_{\text{DP}}\,\hat{\Gamma}_{\text{DP}}(c_0 + \xi_{\text{DO}}\,c_1) \frac{\nabla (c_0 + \xi_{\text{DO}}\,c_1)}{c_0 + \xi_{\text{DO}}\,c_1} \tag{S33}$$

The numerical results presented in the manuscript are obtained by using the parameters shown in Table S4.

**Table S4. Simulation parameters.**

| Parameter | Values |
|---|---|
| $\rho$ | $10^3$ kg/m$^3$ |
| $\eta$ | $0.9 \times 10^{-3}$ Pa s |
| $w$ | 300 µm |
| $h$ | 45 µm |
| $l$ | 5.225 mm |
| $U_0$ | 13.52 mm s$^{-1}$ |
| $c_L$ | 0.01 mM or 1 mM |
| $c_H$ | 1 mM or 100 mM |
| $2a$ | 543 nm or 1222 nm |
| Re | 4.51 |
| Pe$_c$ | 3009 |
| Pe$_n$ | $4.54 \times 10^6$ or $10.22 \times 10^6$ |
| $\xi_{\text{DO}}$ | $2.91 \times 10^{-6}$ or $51.2 \times 10^{-6}$ |
| $\xi_{\text{DP}}$ | from $7.29 \times 10^{-5}$ to $2.02 \times 10^{-4}$ |

## Dynamics of 0.5 µm red fluorescence particles at lower ionic strengths

The weaker focusing of particle under lower ionic strength conditions, evidenced by Fig. 2 in the main manuscript, could be attributed to a reduced diffusiophoresis mobility, which is responsible for the particle migration along the channel depth-wise ($y$) direction and the accumulation at the top and bottom walls of the device. According to our calculations of diffusiophoresis mobility (Figure S3), the ratio $\Gamma_{\text{DP}}(c = 0.5\,\text{mM})/\Gamma_{\text{DP}}(c = 5\,\text{mM})$ is equal to 0.98, 0.82, 0.79, 0.94 and 0.78 for 20 nm, 100 nm, 200 nm, 0.5 µm and 1 µm particles, respectively. The calculated reductions in diffusiophoresis mobility from the average salt concentration under moderate ionic strength conditions (5 mM) to the average salt concentration under lower ionic strength conditions (0.5 mM) vary only from a few percent to twenty percent, depending on the particle type. Such moderate reductions are not sufficient to explain the absence of peak formation at lower ionic strength for sub-micron particles, suggesting that at the lower ionic strength, particle diffusiophoresis mobility may be even lower than estimated.

To further investigate these observations, the dynamics of 0.5 µm red fluorescence particles was simulated under low ionic strength conditions ($c_L = 0.01\,\text{mM}$ and $c_H = 1\,\text{mM}$, compactly expressed as 0.01-1mM in the reminder of the document). Figure S6 shows the simulated transverse profile of the normalized particle concentration $n/n_0$ at a normalized distance from the downstream junction of $z/w_m = 25$ (i.e., $z = 5\,\text{mm}$). The numerical profiles are calculated from the simulated particle concentration field by averaging the concentration over the channel depth ($y$ direction). Figure S7 shows the simulated particle concentration field on the top-right quadrant of the plane perpendicular to the flow direction and located at $z/w_m = 25$. Initially,

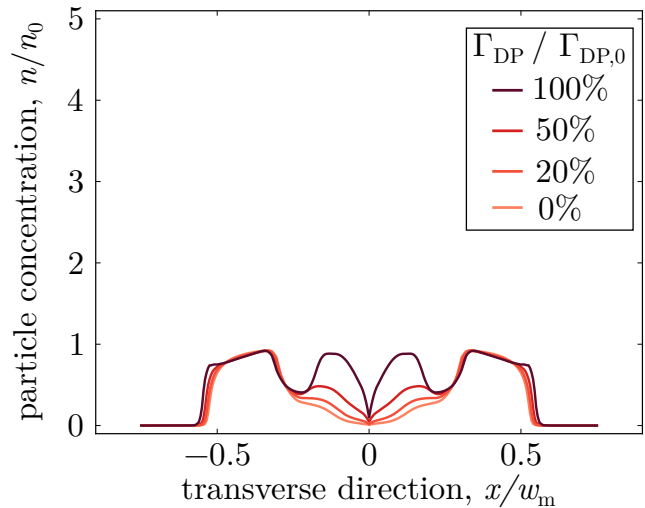

**Fig. S6. Simulated particle concentration profiles at lower ionic strength conditions.** Simulated normalized particle concentration profile $n/n_0$ at a normalized distance from the downstream junction of $z/w_m = 25$ (i.e., $z = 5\,\text{mm}$) for 0.5 µm particles at the lower ionic strength conditions (0.01-10 mM). The value of the diffusiophoresis mobility $\Gamma_{\text{DP,0}}$ in the model is varied from 100% to 0% of the predicted value $\Gamma_{\text{DP,0}}$ calculated according to Eq. S6.

  

the value of particle diffusiophoresis mobility $\Gamma_{DP}$ in the numerical model was set equal to the predicted value $\Gamma_{DP0}$ calculated according to Eq. S6. For this range of salt concentrations, the effective mobility is always negative, hence the in-plane particle velocity is directed inward anywhere near the wall (Figure S7b). Consequently, the simulated particle concentration profile (Figure S6) as well as the experimental fluorescence intensity profile (Figure 2a in the main manuscript) show no peak in the outer regions of the channel. On the other hand, the numerical analysis predicts the formation of inner peaks migrating inwards (Figure S6), but these are not visible in the experimental profile. To shed light on the cause of the observed discrepancy, numerical simulations were performed with a reduced value of particle diffusiophoresis mobility $\Gamma_{DP}$ – either 0%, 20% and 50% of the predicted value $\Gamma_{DP0}$. By comparing the numerical and experimental profiles, it can be concluded that the diffusiophoresis of particles at the lower ionic strength is lower that the one predicted by Eq. S6.

The particle concentration field on the top-right quadrant of the channel cross-section is shown in Figure S7 for varying diffusiophoresis mobility $\Gamma_{DP}$, namely 100%, 50% and 20% and 0% of the predicted value $\Gamma_{DP0}$. As expected, both Figure S6 and Figure S7 show that a smaller diffusiophoresis mobility results in less intense accumulation of colloids. The normalized intensity of the in-plane particle velocity fields are shown in Figure S8 together with the regions of highest colloid concentration (green regions) for varying diffusiophoresis mobility $\Gamma_{DP}$. Note that when the diffusiophoresis mobilities vanishes (Figs. S8d), particles no longer accumulate anywhere in the device. As expected, at the lower ionic strength the migration of colloids towards the top and bottom walls of the channel is slowed down by the reduced value of diffusiophoresis mobility, and the particle accumulation peaks are no longer in close proximity to the walls as it occurs under higher ionic strength conditions (Figure S8e). Consequently, although at lower salt concentrations the total particle velocities at the walls are more intense due to the higher values of effective mobilities, the inward transverse migration of the particle peaks is slower due to the larger distance between the wall and particle accumulation regions.

## Dynamic light scattering analysis for the 0.5 µm and 1 µm red particle separation experiments

To validate further the outcome of the separation experiments for a 0.5 µm and 1 µm red fluorescent particle mixture under high ionic strength conditions (1-100 mM) in the double $\Psi$-junction device depicted in Figure 4 of the main manuscript, both the inlet and extracted central outlet streams of the device were analyzed via dynamic light scattering. An average particle diameter of $563 \pm 3$ nm and $501 \pm 6$ nm were determined for the inlet and central outlet solutions, respectively, highlighting an 11% reduction in the average particle size. These measurements are consistent with the removal of the 1 µm colloids from the mixture. In addition, the polydispersity index (PDI), which quantifies the uniformity of a particle population, showed a significant decrease from $0.083 \pm 0.078$ in the inlet solution to $0.021 \pm 0.008$ in the central outlet solution. According to the ISO standard document 22412:2017 (59), typically the PDI has values less than 0.07 for a monodisperse sample of spherical particles. Values smaller than 0.05 are achieved only for highly monodisperse solutions. Therefore, the measured PDI values demonstrate that the colloidal solution exiting the central outlet was highly monodisperse and its particle size distribution was narrower than the one of the binary mixture injected into the device. This result confirms the successful separation of the smaller particles from the mixture.

## Dynamics of yellow-green and red fluorescent 0.5 µm particles

To interpret the experimental observations reported in Figure 5 of the main manuscript, we calculated the effective mobility profile along the transverse direction $x$ at short ($z/w_m = 45$, i.e. $z = 9$ mm) and long ($z/w_m = 175$, i.e. $z = 35$ mm) distances from the downstream junction for yellow-green and red fluorescent 0.5 µm particles. First, we use the numerical model to determine the dimensionless salt concentration field $c(x, y, z)$ in a long rectangular channel of length $l = 35$ mm $+ 5 \times h = 35.225$ mm. Such a large computational domain requires a large number of elements (ca. $1.3 \times 10^6$). To limit the number of degrees of freedom and, thus, the memory requirement for the simulation, the field $c$ was calculated only up to the $0^{th}$ order, $c_0$. In other words, the effect of the diffusioosmotic flow on the salt concentration field is neglected. This is a reasonable assumption because the $1^{st}$ order term in Eq. S25 is small compared to the $0^{th}$ order. Hence, this approximation will not significantly affect the estimate of the mobility coefficients as they depend only weakly on the salt concentration. The profile of the effective mobility along the transverse direction was then calculated as $\Gamma_{\text{eff}}(x, z) = \Gamma_{\text{DP}}(c_{wall}(x, z)) - \Gamma_{\text{DO}}(c_{wall}(x, z))$, with $c_{wall} = c_H \cdot c(x/w, h/2w, z/w)$ the dimensional salt concentration profile along the transverse direction, located at the top wall of the device and distance $z$ downstream of the junction. Figure S9 shows the effective mobility profiles for both particle species at short and long distances from the junction under high ionic strength (1-100 mM, Figure S9a,b) and moderate ionic strength (0.1-10 mM, Figure S9a,b) conditions.

When high ionic strength conditions were imposed in the double $\Psi$-junction device, the dynamics of yellow-green 0.5 µm fluorescent carboxylate polystyrene particles differed significantly in comparison to the red fluorescent particles at extended distances downstream of the junction (Figure 6). This is because of the opposite zeta potential sensitivities of the two particle species resulting in significantly different effective mobilties at large distance from the junction, as detailed in the main manuscript. To verify further this interpretation, we conducted the same experiments but at a moderate ionic strength condition (0.1-10 mM). The epi-fluorescence micrographs and corresponding fluorescence intensity profiles along the transverse direction

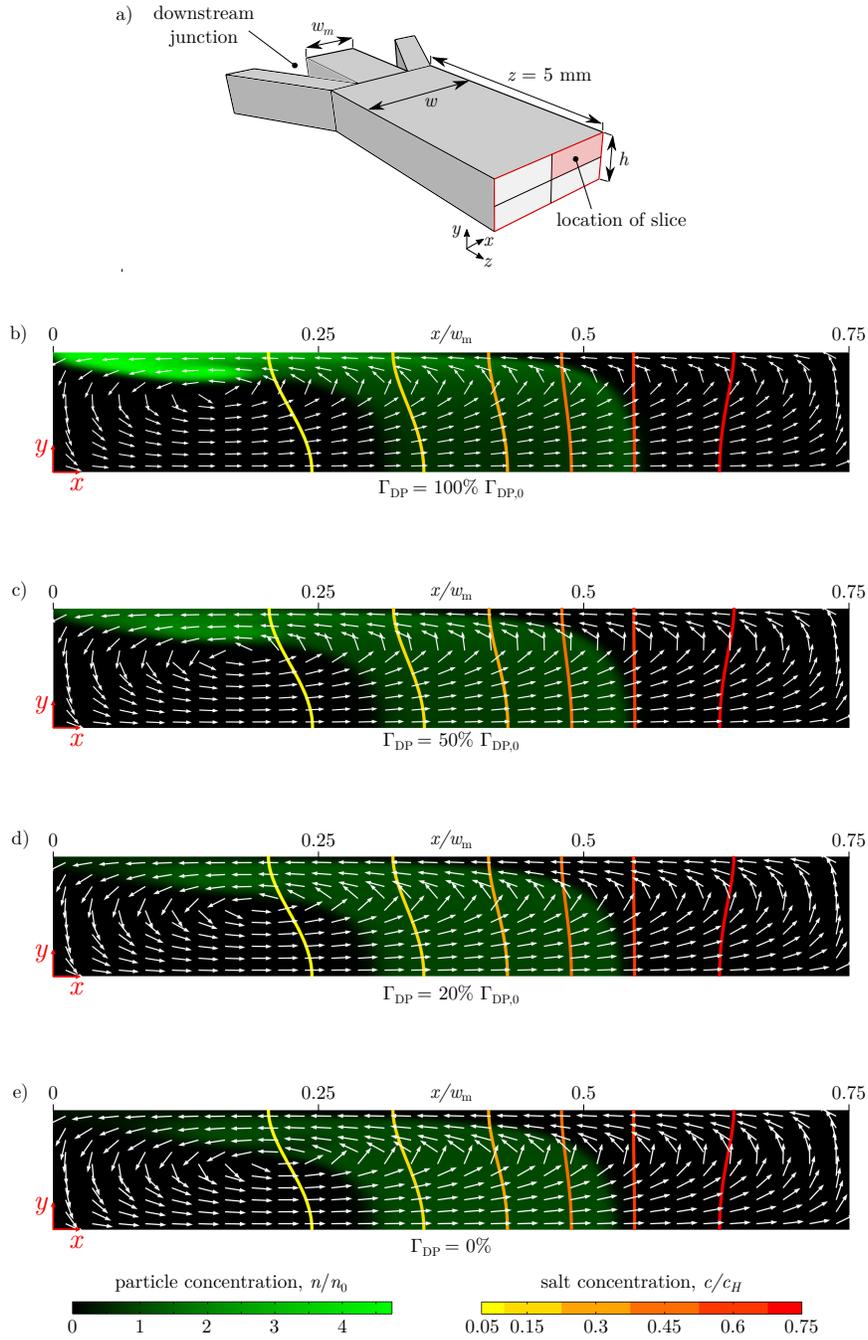

**Fig. S7. Simulated particle concentration distributions at lower ionic strength conditions.** (a) Schematic of the channel in the region near the downstream junction where particle dynamics were simulated. (b-d) Simulated normalized particle distribution (c) on the cross-section slice highlighted in red in panel (a) at $z/w_m = 25$ (i.e., $z = 5$ mm) for $0.5\,\mu$m particles at the lower ionic strength conditions (0.01-10 mM). The value of the diffusiophoresis mobility $\Gamma_{DP,0}$ in the model is varied from $100\%$ to $0\%$ of the predicted value $\Gamma_{DP,0}$ calculated according to Eq. S6. The white arrows represent the streamlines of the total particle velocity $u_p$ and the solid colored lines are the salt concentration isolines.

$x$ are reported in Figure S10 for both particles species and at short (i.e. $z/w_m = 45$) and long (i.e. $z/w_m = 175$) distances from the downstream junction. As suggested by the zeta potential measurements in Figure S2 and the effective mobility estimates in Figure S9, under this salt concentration range, the difference between the zeta potential of the two particle species is not high enough to trigger distinct particle dynamics by reversing the sign of the effective mobility of any particle species in the inner region of the channel ($x/w_m < 0.25$). It follows that throughout the channel both particle species migrate inward and accumulate in the center of the channel, as illustrated in Figure S10. In conclusion, the separation of polystyrene particles with opposite sensitivities of zeta potential to local salt concentration is possible only under selected ionic strength conditions.

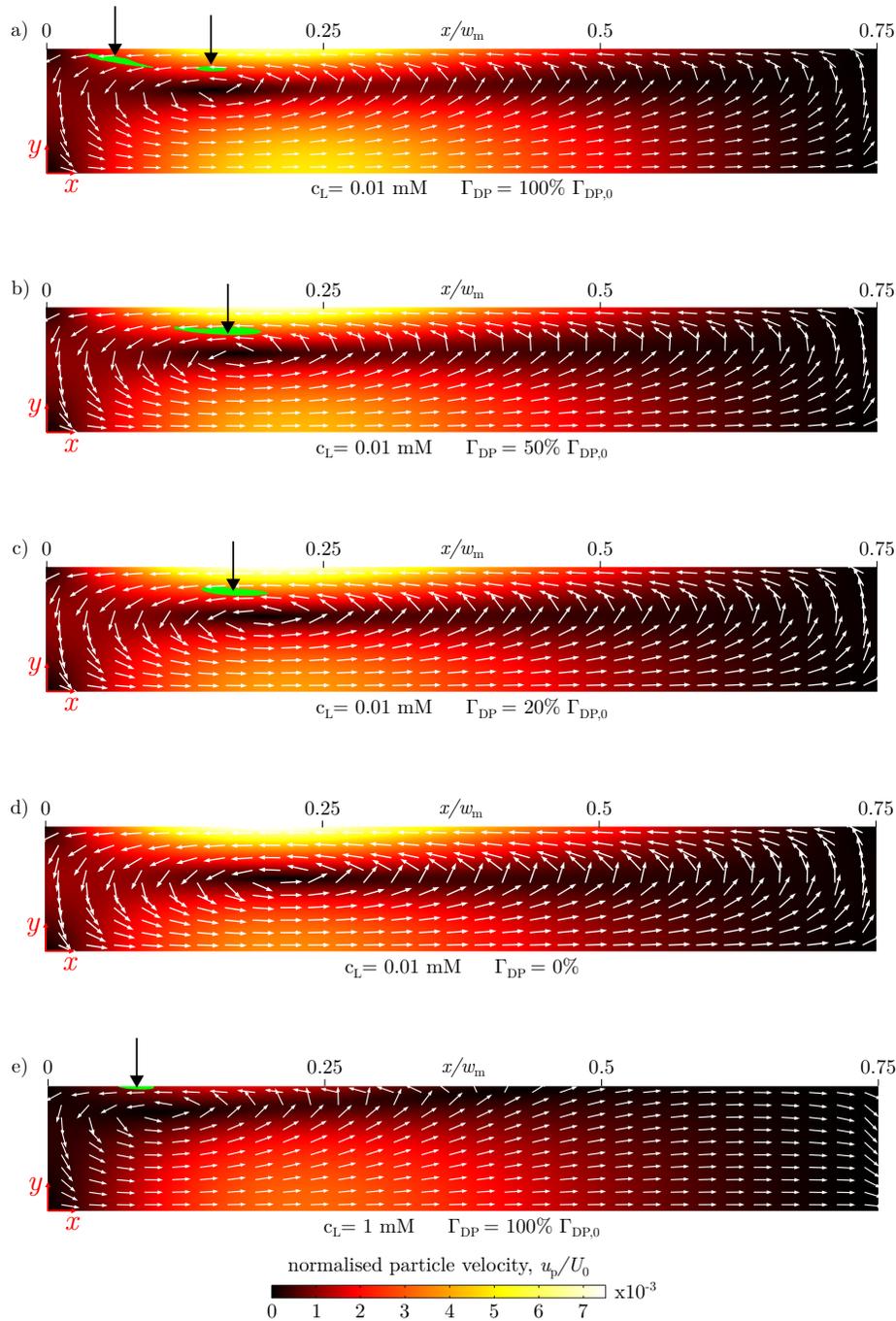

**Fig. S8. Simulated in-plane particle velocity field at lower ionic strength conditions.** Simulated normalized intensity of the in-plane particle velocity on the cross-section slice at $z/w_m = 25$ (i.e., $z = 5$ mm), highlighted in red in Figure S7a. (a-d) Normalized particle velocity intensity for $0.5\,\mu$m particles at the lower ionic strength conditions (0.01-10 mM). The value of the diffusiophoresis mobility $\Gamma_{DP,0}$ in the model is varied from $100\%$ to $0\%$ of the predicted value $\Gamma_{DP,0}$ calculated according to Eq. S6. (e) Normalized particle velocity intensity for $0.5\,\mu$m particles at the higher ionic strength conditions (1-100 mM). The white arrows represent the streamlines of the total particle velocity $u_p$ and the green-shaded areas are the regions with the highest concentration of colloids.

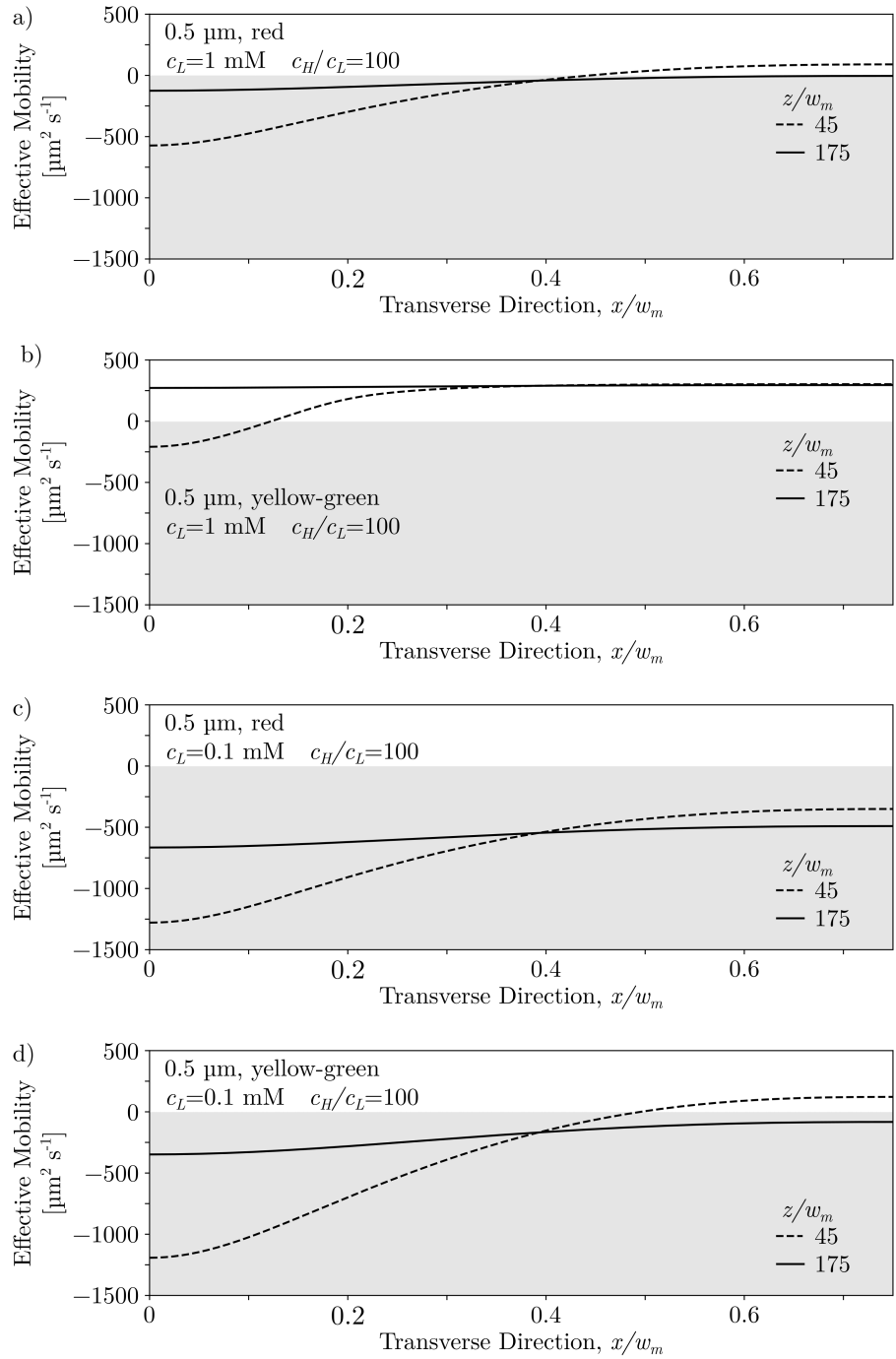

**Fig. S9. Effective mobility profiles along the transverse direction $x$ at the top wall of the channel ($y = h/2$).** Profiles for $0.5\,\mu m$ red (a,c) and $0.5\,\mu m$ yellow-green (b,d) fluorescent particles under high (a,b) and intermediate (c,d) ionic strength conditions.

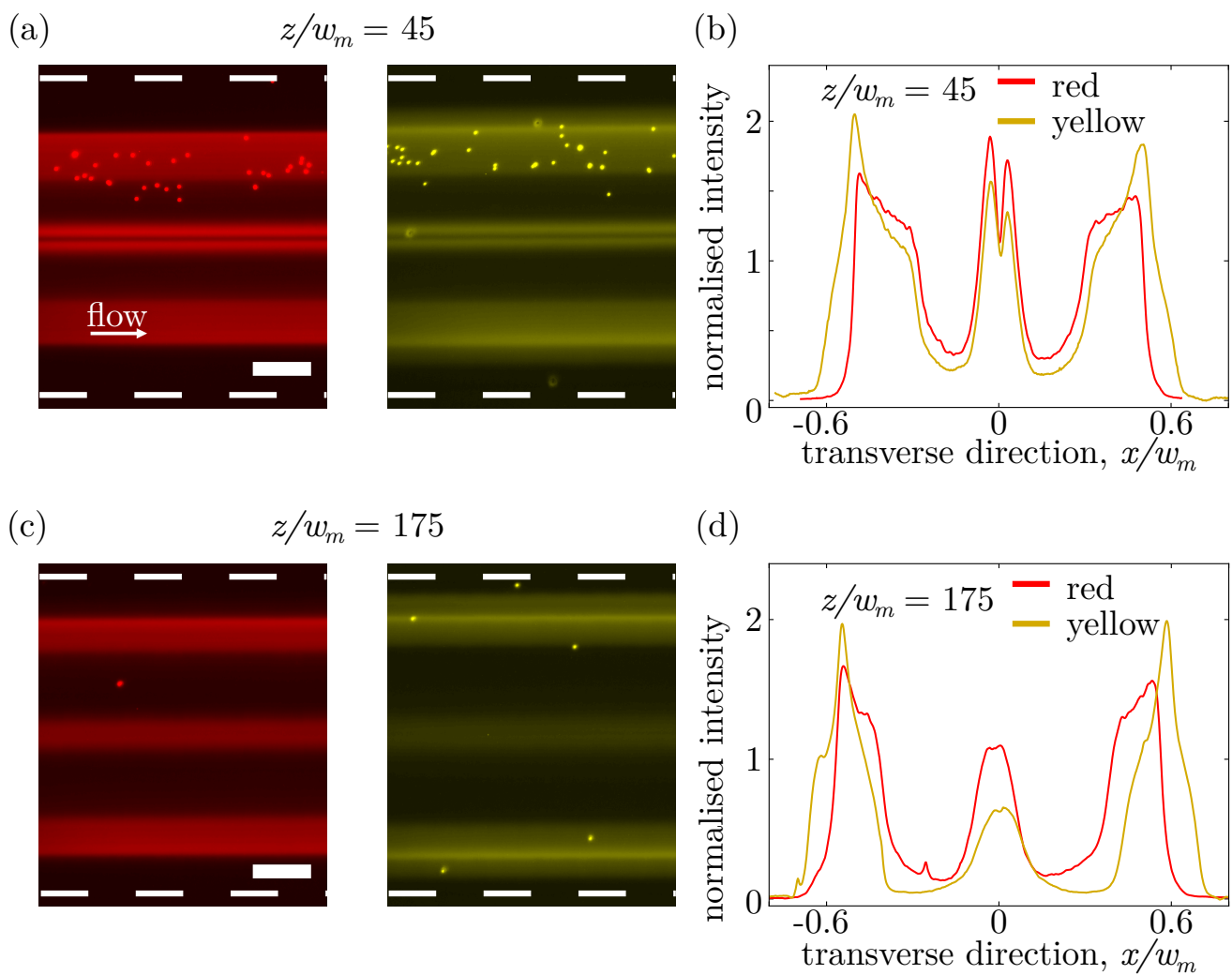

**Fig. S10. Epi-fluorescence micro-graphs and corresponding transverse fluorescence intensity profiles under moderate ionic strength conditions in a long double junction device.** A low salt concentration ($c_L = 0.1$ mM) is injected in all channels of the upstream junction with the side channels containing the binary 1:1 colloid mixture of red-fluorescent and yellow-green fluorescent 0.5 μm carboxylate polystyrene particles. High salt concentration ($c_H = 10$ mM) streams are injected in the side channels of the downstream junction. (a,b) Micro-graphs and fluorescence intensity profiles for red and yellow-green particles at a short distance from the downstream junction ($z/w_m = 45$, i.e. $z = 9$ mm). (c,d) Micro-graphs and fluorescence intensity profiles for red and yellow-green particles at a long distance from the downstream junction ($z/w_m = 175$, i.e. $z = 35$ mm). Scale bar is 50 μm.